\documentclass[aps,prb,twocolumn,showpacs,floatfix,a4paper]{revtex4-2}
\usepackage{graphicx}
\usepackage[intlimits]{amsmath}
\usepackage{amssymb}
\usepackage{xcolor}

\DeclareMathOperator{\rank}{rank}
\usepackage{expl3}
\ExplSyntaxOn
\pdf_version_gset:n{1.7}
\ExplSyntaxOff

\begin{document}

\title{Localization and Flat Bands in  Bond-Inflated Lattices}

\author{Richard Berkovits}
\affiliation{Department of Physics, Jack and Pearl Resnick Institute, Bar-Ilan University, Ramat-Gan 52900, Israel}

\begin{abstract}
We study localization and flat-band formation in lattices generated by repeated  bond inflation of square, honeycomb, and triangular parent lattices. Replacing each bond by a finite tight-binding chain produces several distinct classes of flat bands: chain-induced flat bands at the eigenenergies of the inserted chains, symmetry-protected zero-energy flat bands in bipartite  bond-inflated lattices, and nearly flat junction bands near the spectral edges for sufficiently long chains. We analyze these mechanisms for ordered Lieb-$L$, super$^{L}$honeycomb, and super$^{L}$triangular lattices, and examine their response to bond disorder, site disorder, random magnetic flux, and randomness in the inflation process itself. While bond and site disorder broaden most flat bands, the zero-energy chiral band and the junction-induced flat bands remain robust under certain perturbations. Remarkably, substantial flat-band features also persist in randomly  bond-inflated graphs, even in the absence of translational symmetry. In particular, the number of zero-energy states is found to be well estimated by the matching deficiency $N-2\nu(G)$, indicating that local tree-like structure continues to control the low-energy nullity. These results identify  bond-inflated lattices as a broad class of systems in which geometry alone generates robust localization in both ordered and random settings.
\end{abstract}

\maketitle

\section{Introduction}

Flat bands in tight-binding systems have attracted sustained interest because they provide a direct route from lattice geometry to localization and strongly correlated phenomena. When dispersion is suppressed, kinetic energy is quenched and the low-energy physics is governed by interactions, disorder, and weak perturbations \cite{Leykam2018,Derzhko2015}. A central mechanism underlying flat-band formation is destructive interference: appropriately phased amplitudes cancel on connecting sites, giving rise to compact localized states (CLS)  \cite{Sutherland1986}. This mechanism is responsible for a wide range of phenomena, including flat-band ferromagnetism in Hubbard models \cite{Lieb1989,Mielke1991,Tasaki1992} and numerous experimental realizations in photonic, cold-atom, and engineered electronic systems \cite{Xia2018,Leykam2018}.

Most known flat-band systems are based on periodic decorated lattices, such as the Lieb and kagome lattices, where symmetry and translational invariance play a central role. However, a natural and largely unexplored question is to what extent flat-band physics survives when the underlying lattice is modified in a way that preserves connectivity but relaxes periodicity. In particular, it is not a priori clear whether flat bands, often associated with fine-tuned interference conditions, can persist in systems lacking translational symmetry or even in random graphs.

\begin{figure*}
\centering
    \includegraphics[width=0.8\textwidth]{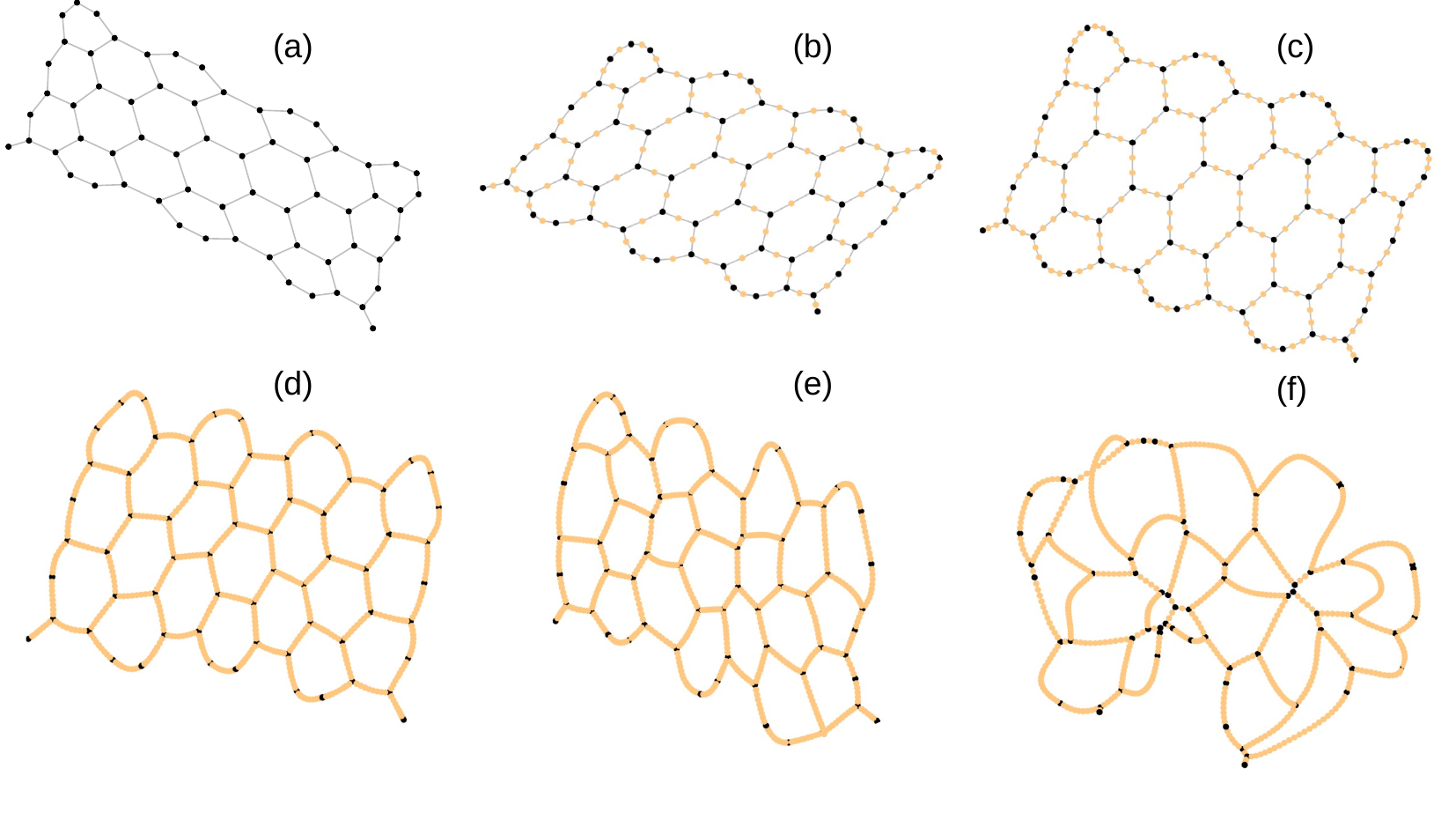}
    \caption{Segments of  bond-inflated honeycomb lattices. For visualization purposes, the  bonds are represented as springs; consequently, distances near the boundaries of the cut-out regions appear distorted. (a) A honeycomb lattice. (b) A superhoneycomb lattice, in which each  bond between the original honeycomb sites (black circles) is inflated into two  bonds connected by a new site (orange circles). (c) A super$^{2}$honeycomb lattice, where each original  bond is inflated twice. (d) A super$^{10}$honeycomb lattice. (e) A honeycomb lattice in which, for each original site, one original  bond is chosen at random and inflated; this random choice is iterated so the averaged length of a chain is ${\bar L}=15$. (f) A honeycomb lattice in which any  bond is chosen at random T Ech step and inflated; again, ${\bar L}=15$ although the fluctuations in the chain length is much larger than for the previous protocol.}
    \label{fig1}
\end{figure*}

In this work we address this question by introducing and analyzing a broad class of  bond-inflated lattices. Starting from a parent lattice, each  bond is replaced by a one-dimensional tight-binding chain. When this procedure is applied uniformly, one obtains periodic lattices such as the Lieb-$L$, super$^{L}$honeycomb and super$^L$triangular families. When the inflation is applied randomly, the resulting structure is a graph with the same large-scale connectivity as the parent lattice but without translational symmetry, and with a broad distribution of chain lengths. This construction therefore provides a unified framework for studying flat-band formation in both ordered and disordered geometries. Examples for such  bond-inflated structures can be seen in Fig. \ref{fig1}.

We show that  bond inflation gives rise to three distinct and physically transparent mechanisms for flat-band formation. First, chain-induced flat bands appear at the eigenenergies of the finite chains replacing the original  bonds and are stabilized by destructive interference at the junction sites. Second, zero-energy flat bands arise in bipartite  bond-inflated lattices due to sublattice imbalance and are protected by chiral symmetry. Third, for sufficiently long chains, junction-induced flat bands emerge near the spectral edges and originate from exponentially localized states centered on high-coordination sites. Although these mechanisms differ in their physical origin, they can all be understood in terms of rank deficiency and the existence of states orthogonal to the coupling subspace linking local motifs to the extended lattice.

A central result of this work is that flat-band features persist well beyond the periodic setting. Even when  bond inflation is implemented randomly, leading to graphs that are neither periodic nor, in general, bipartite, robust spectral signatures of flat bands remain. In particular, we find that a substantial number of zero-energy states survives and is well approximated by the matching-deficiency expression $N-2\nu(G)$, where $\nu(G)$ is the maximum matching number of the underlying graph. This behavior suggests that local tree-like structure and self-averaging over loops play a key role in controlling the low-energy spectrum of these systems.

We further investigate the robustness of the different classes of flat bands under various types of disorder, including bond disorder, site disorder, and random magnetic flux. While most flat bands are broadened by disorder, the zero-energy flat band and the junction-induced flat bands exhibit remarkable stability under appropriate perturbations. Random magnetic flux, in particular, gaps the Dirac cones but leaves the flat bands essentially unaffected, highlighting their structural origin.

The paper is organized as follows. In Sec.~II we formulate the tight-binding Hamiltonian for  bond-inflated lattices in graph-theoretic terms. In Sec.~III we present the different mechanisms for flat-band formation. Sections IV–VI analyze the spectral properties of ordered  bond-inflated lattices, their response to disorder, and the effects of random inflation. In Sec.~VII we discuss the role of matching deficiency in controlling the number of zero-energy states. Finally, we summarize our results and outline possible extensions.

\section{Tight-binding Hamiltonian for  bond-inflated lattices}

A convenient way to formalize an  bond-inflated lattice is through its graph representation $G=(V,E)$, where each lattice site is a vertex $v\in V$ and each allowed hopping process is an edge $(u,v)\in E$. In the single-particle tight-binding description, the Hilbert space is spanned by the site states $\{\lvert v\rangle\}_{v\in V}$, and the Hamiltonian is determined by the weighted adjacency structure of the graph:
\begin{equation}
H = \sum_{v\in V}\varepsilon_v\,|v\rangle\langle v|
+ \sum_{(u,v)\in E}\left(-t_{uv}\,|u\rangle\langle v| - t_{vu}\,|v\rangle\langle u|\right).
\label{eq:H}
\end{equation}
Thus, the connectivity of $G$ fixes which off-diagonal matrix elements are nonzero, while the  bond weights $t_{uv}$ and on-site terms $\varepsilon_v$ specify their values.

In our constructions it is useful to partition the vertices into the set of original sites $O$ and the set of chain sites $C$, so that $V=O\cup C$. We denote the corresponding cardinalities by
\[
N_O = |O|, \qquad N_C = |C|, \qquad N = |V| = N_O + N_C.
\]
 Bonds connect each original junction site $o\in O$ to the first site of each incident chain and connect neighboring sites along each chain; see Fig.~\ref{fig1}. In the uniform case, $t_{uv}=t$ on all edges and $\varepsilon_v=0$, so $H$ is simply $-t$ times the adjacency matrix of $G$.

\section{Mechanisms for flat bands in  bond-inflated lattices}

Flat bands in  bond-inflated lattices, such as Lieb-$L$, super$^{L}$honeycomb and super$^L$triangular systems, can be understood from several complementary theoretical viewpoints. Although these approaches differ in language and emphasis, they ultimately describe different manifestations of rank deficiency or destructive interference in the tight-binding Hamiltonian.

For bipartite nearest-neighbor tight-binding models with no intra-sublattice hopping, the Bloch Hamiltonian can be written in block off-diagonal form,
\begin{equation}
H(\mathbf{k})=
\begin{pmatrix}
0 & B(\mathbf{k})\\
B^\dagger(\mathbf{k}) & 0
\end{pmatrix}.
\end{equation}
If the number of sites on the two sublattices differs, $N_A \neq N_B$, the rank-nullity theorem guarantees at least $|N_A-N_B|$ zero eigenvalues, producing symmetry-protected flat bands at $\varepsilon=0$ \cite{Inui1994,Lieb1989,Mielke1991,Tasaki1992}. In  bond-inflated Lieb lattices, zero-energy flat bands arise whenever  bond inflation preserves bipartiteness while introducing sublattice imbalance.

Flat bands can also be constructed explicitly in real space by building compact localized states whose amplitudes cancel at all junctions that connect to the rest of the lattice \cite{Inui1994,Bergman2008,Flach2014,Leykam2018}. For zero-energy bands, cancellation typically occurs on one sublattice due to chiral symmetry. For nonzero-energy flat bands in  bond-inflated systems, cancellation occurs at the junctions between inflated chains and the original sites. The Schr\"odinger equation at a junction $o\in O$ reads
\begin{equation}
\sum_{c\ni o} -t\,\psi^{(c)}_{1} = \varepsilon\,\phi_o,
\label{eq:interference}
\end{equation}
where $\phi_o$ is the wave-function amplitude on the original site $o$, and $\psi^{(c)}_j$ is the amplitude on site $j$ of chain $c$, with $j=1$ denoting the site closest to $o$. At $\varepsilon=0$, Eq.~(\ref{eq:interference}) reduces to
\begin{equation}
\sum_{c\ni o} -t\,\psi^{(c)}_{1} = 0
\qquad \forall o\in O.
\end{equation}
Ensuring that the amplitude at each original site vanishes and preventing hybridization with neighboring chains yields $|N_O-N_{\mathrm{ch}}|$ independent solutions, where $N_{\mathrm{ch}}$ denotes the number of inflated chains in the lattice.

Nonzero-energy flat bands may arise when each original  bond is replaced by a finite tight-binding chain of length $L$. Such a chain supports $L$ discrete eigenmodes with eigenvalues
\begin{equation}
\varepsilon_m = -2t\cos\!\left(\frac{m\pi}{L+1}\right),
\qquad m=1,\dots,L.
\label{eq:chain}
\end{equation}
Superpositions of these chain modes can satisfy the junction interference condition, thereby forming flat bands at nonzero energies. This mechanism has been demonstrated explicitly in extended Lieb and Lieb-$L$ lattices \cite{Zhang2017,Hanafi2022}.

Some flat-band lattices, including Lieb and kagome lattices, can be interpreted as line graphs or partial line-graph constructions \cite{Mielke1991,Bergman2008}; see Appendix~\ref{app_line}. A standard example is the kagome lattice, which is the line graph of the honeycomb lattice. In such cases, the Hamiltonian admits a factorized form $H=B^\dagger B$, so flat bands correspond to the kernel of $B$. Decorated extensions enlarge this kernel or create block-diagonal sectors that remain dispersionless.

Flat bands may also be interpreted as antisymmetric bound states that decouple from dispersive continua through destructive interference, analogous to Fano lattice constructions \cite{Flach2014}. In  bond-inflated Lieb systems, localized chain modes can decouple from the original sites because of symmetry-enforced orthogonality.

A different route to nonzero-energy flat bands arises from junction states. Physically, these states arise from the enhanced kinetic energy associated with a larger number of nearest neighbors at a junction \cite{khatua21,he23}. As detailed in Appendix~\ref{app_junction}, when bonds are replaced by infinitely long chains in a lattice whose original coordination number is $k$, a localized junction state appears with energy
\begin{equation}
\varepsilon = \pm t\sqrt{\frac{k^2}{k-1}},
\label{eq:junc_e}
\end{equation}
and eigenvectors
\begin{equation}
\psi^{(c)}_j = (\pm 1)^j e^{-j/\zeta}\phi_o,
\label{eq:junc_p}
\end{equation}
for each chain $c=1,\dots,k$. Here $\phi_o$ is fixed by normalization, $j$ labels the chain sites starting from the junction, $\zeta = 2/\ln(k-1)$, and the $\pm$ sign corresponds to the ground and highest excited states, respectively. Such a localized state exists at every original site $o\in O$.

These solutions correspond to energies lying below and above the one-dimensional chain band. The associated localization length is $\zeta=O(1)$. Consequently, for $L\gg \zeta$, the overlap between eigenvectors centered at neighboring original sites is of order $e^{-L/\zeta}$, leading to a band width proportional to $e^{-L/\zeta}$. Thus, when $L\gg \zeta$, the junction band becomes exponentially narrow. Moreover, as long as there are sufficiently many original sites for which all attached chains are much longer than $\zeta$, although not necessarily of equal length, the band remains robust.

To summarize, all of these viewpoints reflect the existence of eigenvectors orthogonal to the coupling subspace that connects local motifs to extended states. Zero-energy flat bands arise from global chiral constraints, whereas finite-energy flat bands arise either from locally confined eigenvectors whose junction interference eliminates momentum-dependent hybridization, or from kinetic-energy gain at the junction. In the following sections, we discuss three specific families of  bond-inflated lattices: Lieb-$L$, super$^{L}$honeycomb, and super$^{L}$triangular.

\section{Ordered  bond-inflated lattices}

In this section we examine the emergence of flat bands in several classes of two-dimensional  bond-inflated lattices.  Periodic boundary conditions are applied throughout all subsequent discussions.

\subsection{Lieb-$L$ lattices}

\begin{figure}
    \includegraphics[width=0.5\textwidth]{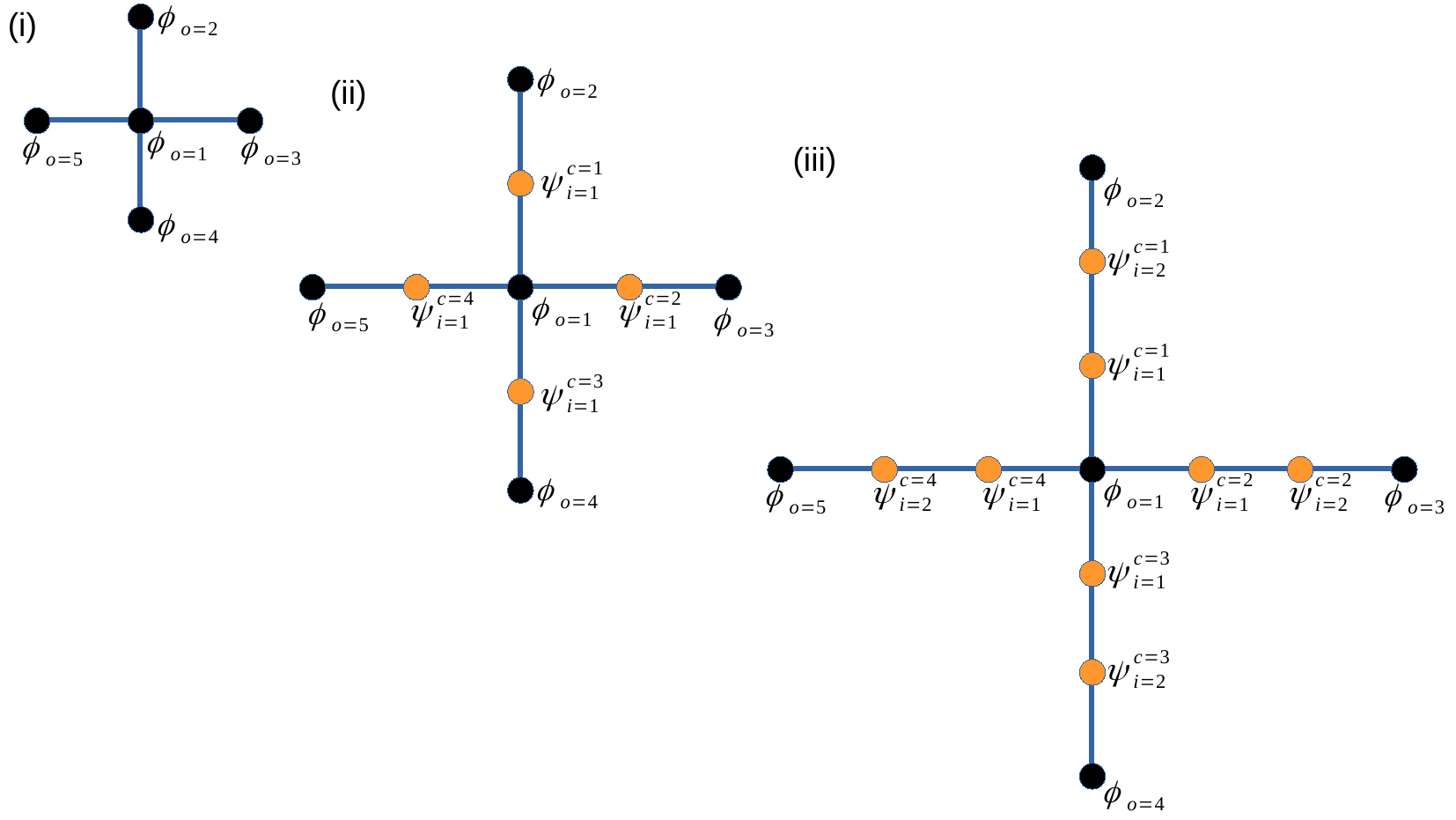}
    \caption{
    Illustration of the  bond-inflation procedure for the Lieb-$L$ lattice. 
(i) A motif extracted from a square lattice, consisting of a central site (black circle) connected to four neighboring sites. 
(ii) Each  bond in (i) is inflated into a new site (orange circle) connected by two  bonds. Applied to the full square lattice, this yields the Lieb-$1$ lattice, i.e., the conventional Lieb lattice. 
(iii) A double inflation step, in which each  bond in (i) is replaced by a chain of length $L=2$. Applied to the square lattice, this produces the Lieb-$2$ lattice.}
    \label{fig2}
\end{figure}

The Lieb lattice and its extensions are canonical systems for studying flat-band physics and geometry-induced localization, both theoretically and experimentally. We begin with the Lieb-$L$ family, obtained from the square lattice by replacing each  bond with a one-dimensional tight-binding chain of length $L$. The inflation procedure is illustrated in Fig.~\ref{fig2}.

\begin{figure}
    \includegraphics[width=0.5\textwidth]{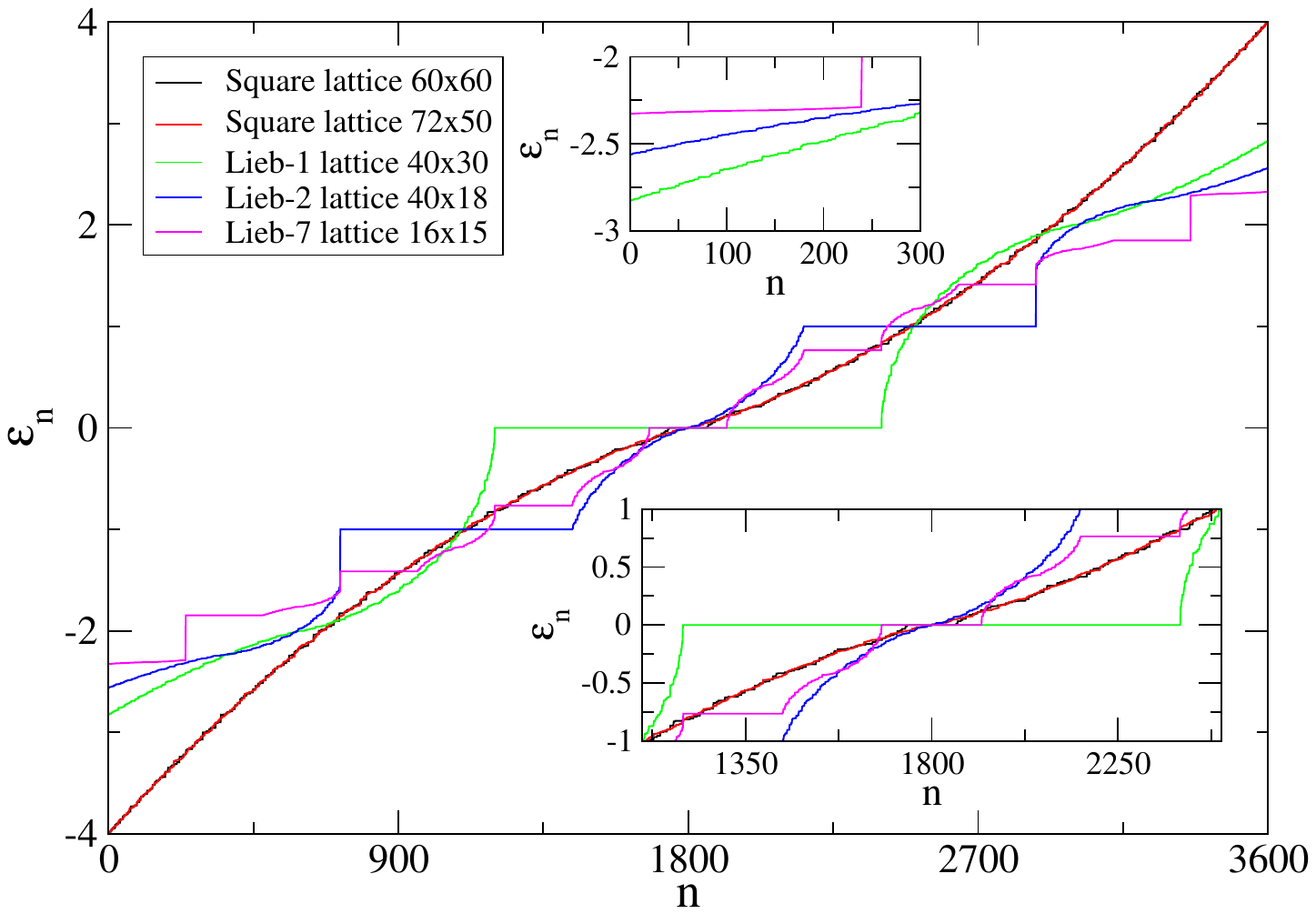}
    \caption{The $n$-th energy level $\varepsilon_n$ as a function of the level index $n$ for square, Lieb, and higher-order Lieb-$L$ lattices, all with a fixed total number of sites $N=3600$ and periodic boundary conditions. Bottom inset: zoom into the zero-energy region. Top inset: zoom into the region below $\varepsilon=-2$. Flat bands appear at different energies, with their degeneracies depending on the value of $L$.}
    \label{fig3}
\end{figure}

It is important to note that accidental degeneracies may occur even in the square lattice itself. This is illustrated in Fig.~\ref{fig3}, where all systems contain a total of $3600$ sites. Regular two-dimensional square lattices with $N_x=N_y=60$ exhibit a small accidental degeneracy at $\varepsilon=0$, as shown in the lower inset of Fig.~\ref{fig3}. In contrast, for $N_x=72$ and $N_y=50$, no such accidental degeneracies occur, and the curve of $\varepsilon_n$ as a function of level index $n$ is smooth.

The original Lieb lattice, i.e., the Lieb-$1$ lattice, is obtained by inflating each bond of the square lattice once, resulting in an  bond-centered square lattice with three sites per unit cell. This lattice hosts a symmetry-protected flat band at zero energy, arising from bipartite sublattice imbalance and destructive interference between hopping paths. In this case $N_C=2N_O$, and therefore the zero-energy flat band contains $N_O$ states.

In Fig.~\ref{fig3}, this corresponds to $N_O=1200$ and $N=N_O+N_C=3600$, where the parent square lattice is arranged with $N_x=40$ and $N_y=30$ sites. A prominent flat band of width $N_O=1200$ dominates the vicinity of $\varepsilon=0$. Another hallmark feature is the appearance of Dirac cones touching the flat band. Dirac cones display the characteristic behavior
\[
\varepsilon_n-\varepsilon_m = \pm v|\mathbf{q}| \approx \varepsilon_m \pm A\sqrt{n},
\]
where $\mathbf{q}$ is the momentum measured from the cone and $A$ is a constant. The coexistence of Dirac cones and a flat band is one of the defining spectral properties of the Lieb lattice and underlies many of its unusual transport, optical, and topological features \cite{Apaja2010,Weeks2010,Shen2010,Nita2013,Slot2017,Diebel2016}. Experimental realizations using scanning tunneling microscopy have confirmed the presence of these flat-band states in artificial electronic lattices \cite{Slot2017}, photonic lattices \cite{Xia2018,Danieli2024}, and compact localized spin-wave excitations in extended Lieb geometries \cite{Krawczyk2023}.

Bond-inflated Lieb lattices introduce additional degrees of freedom into the unit cell and lead to multiple flat bands at both zero and nonzero energies \cite{Bhattacharya2019}. A transparent way to understand this is to treat each inflated chain as a local object with discrete energies $\varepsilon_m$ given by Eq.~(\ref{eq:chain}). For each chain eigenmode, the junction interference condition becomes
\begin{equation}
\sum_{c\ni o} -t\,\psi^{(c,m)}_{1}(\varepsilon_m) = \varepsilon_m \phi_o(\varepsilon_m),
\qquad \forall o\in O,
\label{eq:interference1}
\end{equation}
where $\psi^{(c,m)}_{1}$ is the amplitude of the $m$-th eigenmode of chain $c$ on the site nearest to the junction. Enforcing vanishing amplitude on the original sites again yields flat-band states localized on the chains.

This behavior is illustrated in Fig.~\ref{fig3}, where the cases of Lieb-$2$ and Lieb-$7$ lattices are shown. For the Lieb-$2$ lattice, the chain spectrum contains two eigenenergies, $\varepsilon=\pm 1$. For $L=7$, the chain spectrum is
\[
\varepsilon=\pm \sqrt{2+\sqrt{2}},\ \pm \sqrt{2},\ \pm \sqrt{2-\sqrt{2}},\ 0,
\]
which matches the energies of the flat bands visible in the figure. Since the total number of sites is fixed at
\[
N=N_O+2L N_O = (2L+1)N_O = 3600,
\]
it follows that
\[
N_O=\frac{N}{2L+1},
\]
which determines the number of states in each flat band, since $N_{\mathrm{ch}}=2N_{O}$, resulting in $|N_{O}-N_{\mathrm{ch}}|=N_{O}$.

The Lieb-$L$ lattices also exhibit a characteristic Dirac-cone structure. For odd $L$, the zero-energy flat band touches two Dirac cones and displays the familiar pseudospin-$1$ behavior. In contrast, the nonzero-energy flat bands, which constitute all flat bands when $L$ is even, are each connected to a single Dirac cone. Such band crossings are commonly described as pseudospin-$1$ Dirac intersections at a time-reversal-invariant momentum (TRIM) \cite{Weeks2010,Shen2010,Nita2013,Diebel2016}.

For larger values of $L$, an additional nearly flat band appears near the spectral edges, as seen in Fig.~\ref{fig3} for $L=7$. The origin of this band, located at $\varepsilon=\pm\sqrt{16/3}\approx \pm 2.31$, lies in the junction-localized states of Eqs.~(\ref{eq:junc_e}) and (\ref{eq:junc_p}).  These states have a large gap below or above the main band ($-2<\epsilon<2$) and can hybridize only with other junction states. Since these states fall of exponentially on a scale $\zeta$, thus if $L \gg \zeta$ the hybridization between the junction states is exponentially small and the junction states band is essentially flat. For the Lieb-$L$ lattice, one expects a localization length $\zeta=2/\ln(3)\approx 1.8$. Consequently, a broadening of order $10^{-3}$ is expected when
\[
\xi \approx -\zeta\ln(10^{-3}) \approx 12.4,
\]
which will be termed the broadening localzation length, consistent with the behavior observed in the top inset of Fig.~\ref{fig3}, where a small broadening remains for $L=7$. 
  
\subsection{Super$^{L}$honeycomb lattices}

The honeycomb lattice and its extensions lie at the center of many research areas, ranging from graphene and other Dirac materials to topological insulators, artificial photonic lattices, cold atoms, and strongly correlated flat-band systems. The honeycomb lattice provides the simplest realization of Dirac fermions in a crystalline system. Its  bond-inflated extensions exhibit a rich variety of phenomena, including topological band structures, flat bands, and unconventional localization effects \cite{CastroNeto2009,Hasan2010,Qi2011,Polini2013,Leykam2018}.

\begin{figure}
    \includegraphics[width=0.5\textwidth]{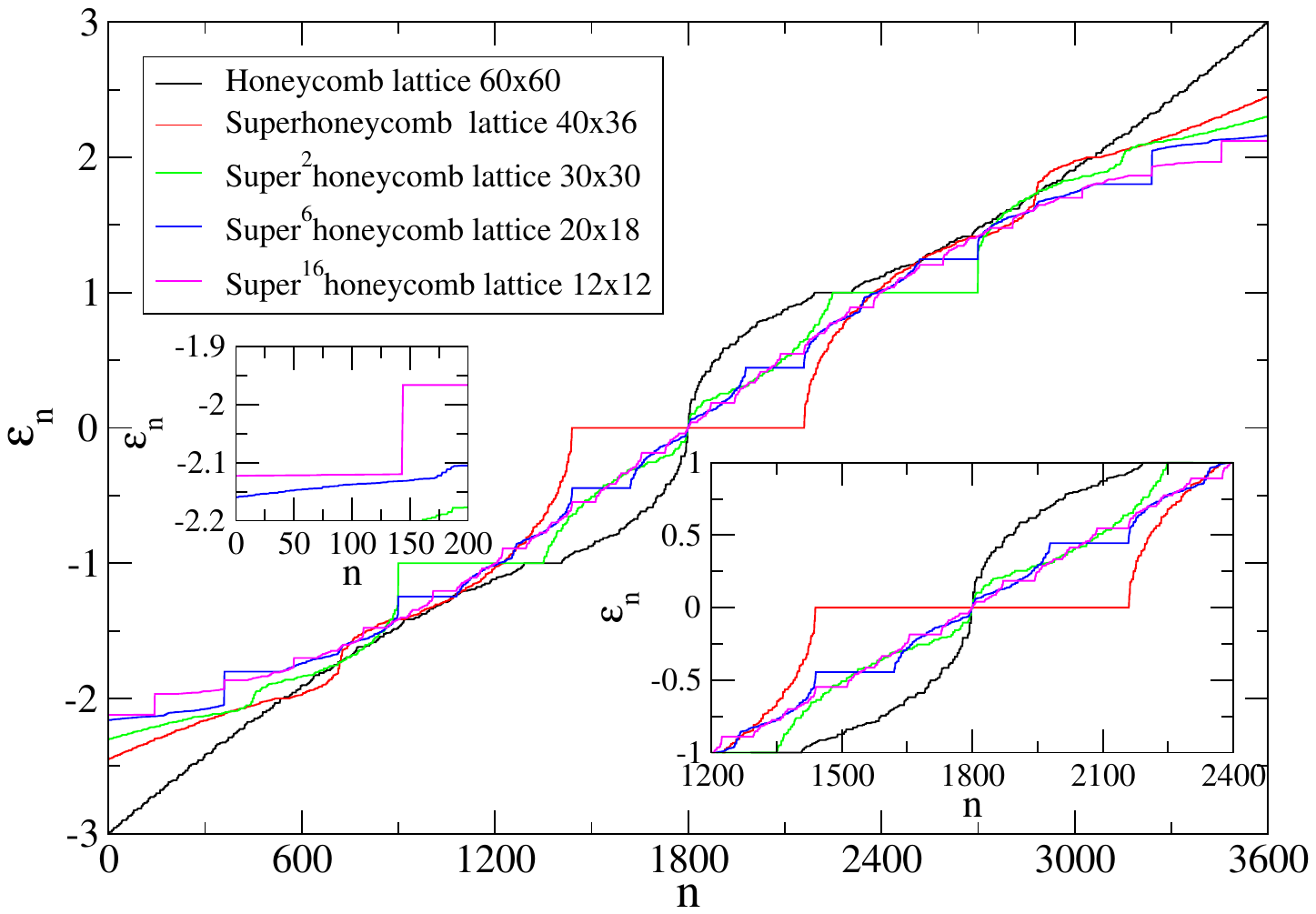}
    \caption{The $n$-th energy level $\varepsilon_n$ as a function of the level index $n$ for honeycomb, superhoneycomb, and higher-order super$^{L}$honeycomb lattices, all with a fixed total number of sites $N=3600$  and periodic boundary conditions. Right inset: zoom into the zero-energy region, where for $L=1$ a highly degenerate flat band appears, while for even values of $L$ a pair of Dirac cones are observed. Left inset: zoom into the region below $\varepsilon=-2$, where for larger values of $L$ (e.g., $L=16$) an additional flat band emerges.}
    \label{fig4}
\end{figure}

Figure~\ref{fig4} shows $\varepsilon_n$ as a function of $n$ for several super$^{L}$honeycomb lattices, each containing $N=3600$ sites. As in the Lieb-$L$ family, flat bands and Dirac cones are clearly visible. The unit cell of the super$^{L}$honeycomb lattice contains two original sites and three inflated  bonds, each replaced by a chain of length $L$. Hence
\[
N = N_O + \frac{3L}{2}N_O,
\]
so for the systems shown in Fig.~\ref{fig4},
\[
N_O = \frac{N}{\frac{3L}{2}+1}.
\]
The number of states per flat band is $|N_{\mathrm{ch}}-N_O|$, where in this family, $N_{\mathrm{ch}}=3 N_O/2$ which gives 
\[
|N_{\mathrm{ch}}-N_O|=\frac{N}{3L+2}
\]
states in each of the $L$ chain-induced flat bands. Their energies are again given by the chain eigenvalues in Eq.~(\ref{eq:chain}).

Unlike the Lieb-$L$ lattices, Dirac cones in the super$^{L}$honeycomb lattice appear both between flat bands and at flat-band energies. As in the Lieb family, the flat-bands touch Dirac-cone structure. The zero-energy flat band touches two Dirac cones  (pseudospin-$1$). The nonzero-energy flat bands tend to connected to a single Dirac cone (pseudospin-$1$, TRIM).


As in the Lieb-$L$ lattice, the super$^{L}$honeycomb lattices also exhibit an additional nearly flat band near the spectral edges for sufficiently large values of $L$; see the upper inset of Fig.~\ref{fig4} for $L=11$. In this case $k=3$, which yields a localized band at
\[
\varepsilon=\pm\sqrt{9/2}\approx \pm 2.12,
\]
according to Eqs.~(\ref{eq:junc_e}) and (\ref{eq:junc_p}). This energy lies closer to the one-dimensional chain band than in the Lieb-$L$ case. The localization length is $\zeta=2/\ln(2)\approx 2.9$, so a band broadening of order $10^{-3}$ is expected when
\[
\xi \approx -\zeta\ln(10^{-3}) \approx 15.9,
\]
consistent with Fig.~\ref{fig4}.

\subsection{Super$^{L}$triangular lattices}

The triangular lattice differs from the lattices discussed above in that it is not bipartite and therefore does not possess particle-hole symmetry. Nevertheless, when its  bonds are inflated, bipartite structure may emerge for odd $L$, and particle-hole symmetry is then restored. Consequently, many of the flat-band features present in the previous lattices also appear here.

\begin{figure}
    \includegraphics[width=0.5\textwidth]{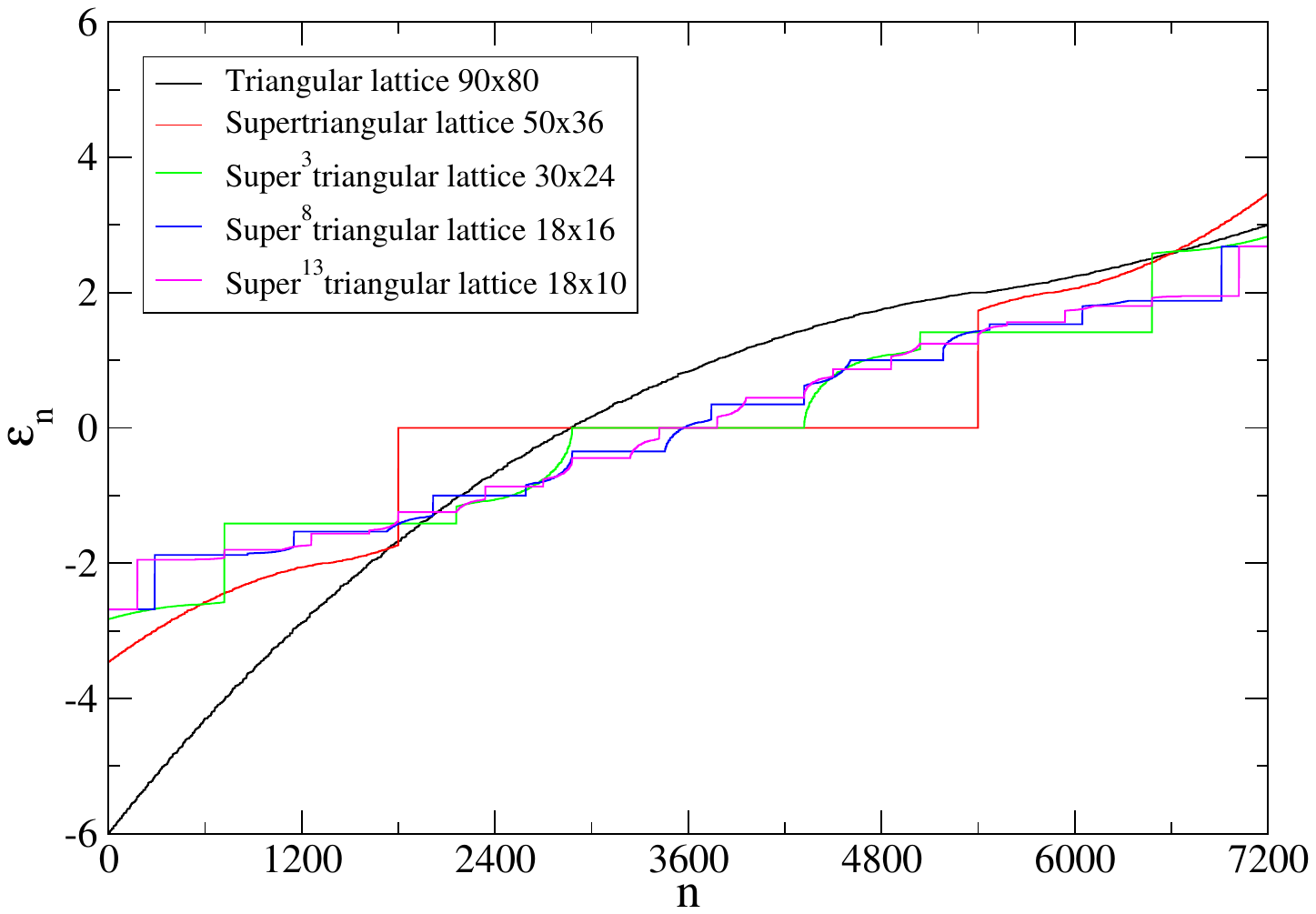}
    \caption{The $n$-th energy level $\varepsilon_n$ as a function of the level index $n$ for triangular, supertriangular, and higher-order super$^{L}$triangular lattices, all with a fixed total number of sites $N=7200$  and periodic boundary conditions. The triangular lattice spectrum lacks particle–hole symmetry, a feature that persists for even values of $L$. In contrast, for odd values of $L$ the system becomes bipartite, exhibits particle–hole symmetry, and supports a flat band at $\varepsilon=0$. Owing to the high coordination number of the original vertices, the flat bands are relatively broad.}
    \label{fig5}
\end{figure}

Figure~\ref{fig5} shows $\varepsilon_n$ as a function of $n$ for several lattices, each containing $N=7200$ sites. The triangular lattice itself is not bipartite and therefore lacks particle-hole symmetry, as is evident from the asymmetric dependence of $\varepsilon_n$ on $n$.

For the super$^{L}$triangular lattice, which to our knowledge has not been analyzed systematically in the literature, there is an important distinction between even and odd values of $L$. For odd $L$, the inflated lattice is bipartite and therefore exhibits particle-hole symmetry. For even $L$, it is not bipartite and no such symmetry is present.

As discussed above, replacing each original  bond by a chain of length $L$ generates $L$ flat bands, corresponding to the $L$ eigenmodes of the isolated chain. Since between any pair of original sites there is only a single chain, each chain eigenvalue contributes one set of states, and the degeneracy of each flat band remains proportional to the mismatch between original and chain degrees of freedom. Thus, in the triangular family,  bond inflation produces $L$ flat bands irrespective of the parity of $L$.

The unit cell of the super$^{L}$triangular lattice contains one original site and three inflated  bonds. Hence
\[
N = N_O + 3L N_O = (3L+1)N_O,
\]
so for the systems in Fig.~\ref{fig5},
\[
N_O = \frac{N}{3L+1}.
\]
The number of states per flat band is
\[
|N_{\mathrm{ch}}-N_O| = 2N_O = \frac{2N}{3L+1},
\]
and the flat-band energies are again given by Eq.~(\ref{eq:chain}).

The Dirac-cone structure is richer than in the previous cases. Depending on $L$, flat bands may be intersected either by a single cone or by a double cone, while some flat bands may not be intersected at all. For example, as seen in Fig.~\ref{fig5}, there are no Dirac cones for $L=1$, while for $L=3$ a double Dirac cone intersects the central flat band, with no additional cones at $\varepsilon=\pm1$.

The super$^{L}$triangular lattice also exhibits an additional nearly flat band near the spectral edges, even more prominently than in the Lieb-$L$ and super$^{L}$honeycomb families. This behavior arises from the larger coordination number, $k=6$, which yields a localized band at
\[
\varepsilon=\pm\sqrt{36/5}\approx \pm2.68
\]
according to Eqs.~(\ref{eq:junc_e}) and (\ref{eq:junc_p}). The corresponding localization length is
\[
\zeta = \frac{2}{\ln 5} \approx 1.24.
\]
A band broadening of order $10^{-3}$ is therefore expected when
\[
\xi \approx -\zeta\ln(10^{-3}) \approx 8.6.
\]
Indeed, as observed in Fig.~\ref{fig5}, the junction band is already nearly flat for $L=8$.

\section{Disorder in geometric ordered  bond-inflated lattices}

We now address whether the flat bands found above remain robust in the presence of disorder. We consider three distinct classes: bond disorder, site disorder, and random magnetic flux  \cite{vidal1998,vidal2001,chalker2010,leykam2013,Danieli2015,Leykam2017,Shukla2019,Cadez2021,Liu2021,Liu2022,Kim2023,Marques2023,Danieli2024a,Rosen2025,Dresselhaus2025}.

\subsection{Bond disorder}

\begin{figure}
    \includegraphics[width=0.5\textwidth]{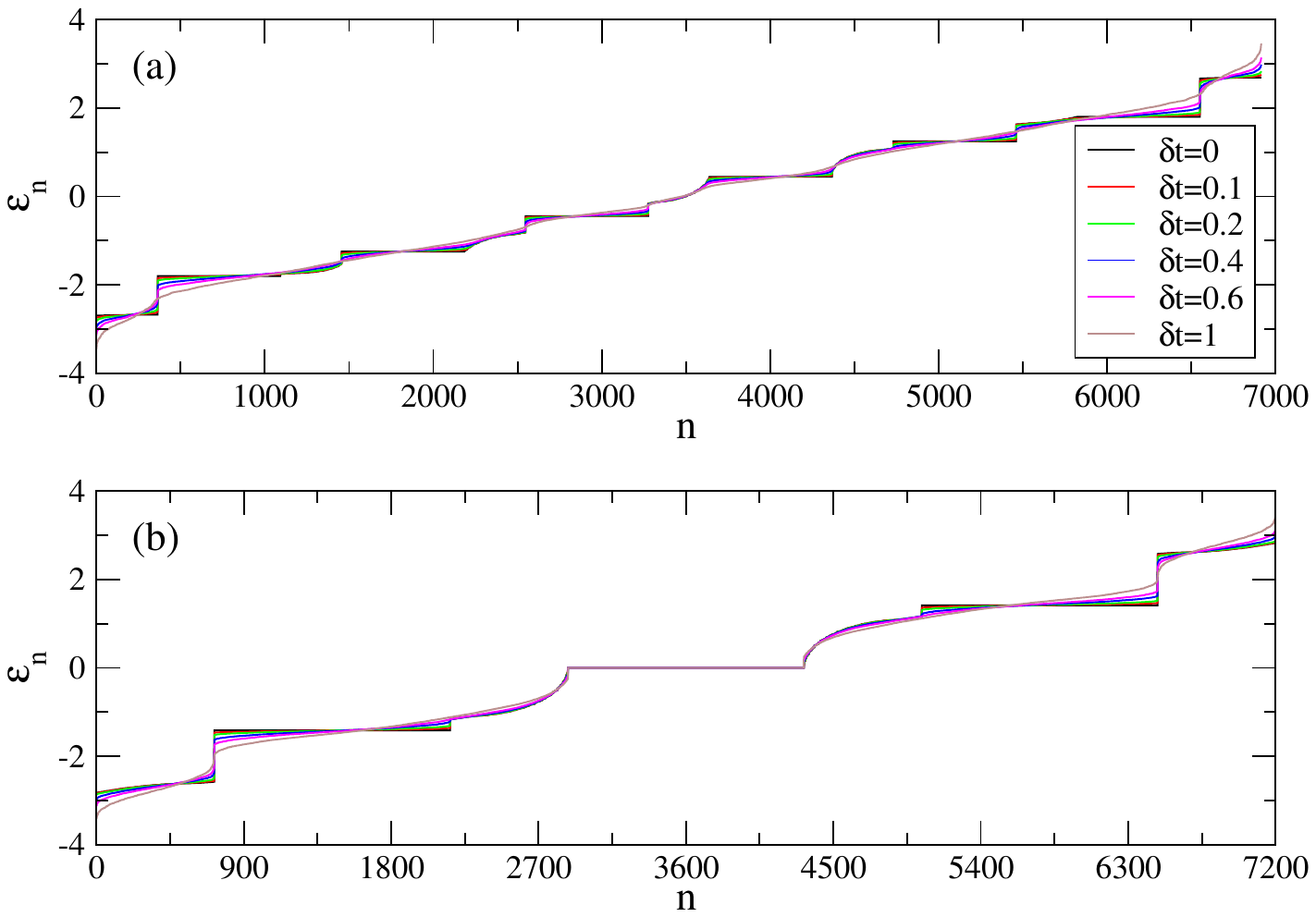}
    \caption{The $n$-th energy level $\varepsilon_n$ as a function of level index $n$ for super$^{L}$triangular lattices with bond disorder. (a) A single realization of a super$^{6}$triangular lattice generated from a $26\times14$ triangular parent lattice, with total size $N=6916$. The hopping amplitudes are independently drawn from a box distribution of width $\delta t$. (b) Same type of disorder for a super$^{3}$triangular lattice generated from a $30\times24$ triangular parent lattice, with total size $N=7200$. All flat-bands except for the one at $\varepsilon=0$ are broadened by disorder.}
    \label{fig6}
\end{figure}

For bond disorder, each hopping amplitude $t_{uv}$ in Eq.~(\ref{eq:H}) is chosen independently from a box distribution of width $\delta t$. As shown in Fig.~\ref{fig6}, bond disorder broadens the flat bands in all cases except for the $\varepsilon=0$ band, which remains unaffected. The robustness of the zero-energy band to bond disorder is a general property of bipartite graphs with chiral symmetry \cite{Inui1994}.

The same qualitative behavior is found for the Lieb-$L$ and super$^{L}$honeycomb lattices. More generally, for all families considered here, the effect of bond disorder depends on the parity of $L$. If $L$ is even, no flat band occurs at $\varepsilon=0$, and all flat bands are broadened. If $L$ is odd, a flat band exists at $\varepsilon=0$ and remains unaffected by bond disorder, whereas all other flat bands broaden.

\subsection{Site disorder}

\begin{figure}
    \includegraphics[width=0.5\textwidth]{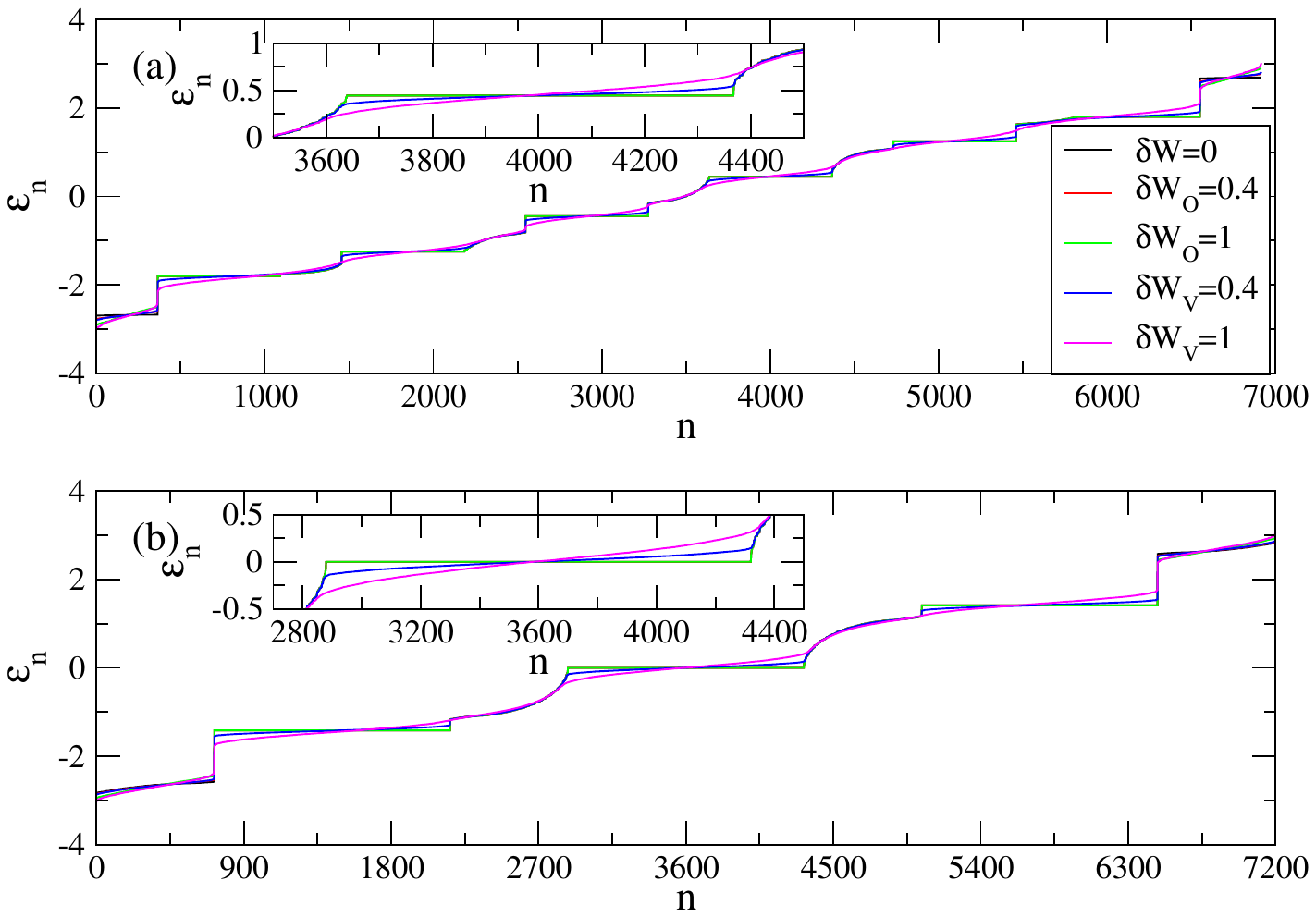}
    \caption{The $n$-th energy level $\varepsilon_n$ as a function of level index $n$ for super$^{L}$triangular lattices with site disorder. (a) A single realization of a super$^{6}$triangular lattice generated from a $26\times14$ triangular parent lattice, with total size $N=6916$. Site energies are drawn independently from a box distribution. Disorder is applied either only to the original sites, with width $\delta W_O$, or to all sites, with width $\delta W_V$. Inset: zoom into a flat band near the center of the spectrum. (b) Same type of disorder for a super$^{3}$triangular lattice generated from a $30\times24$ triangular parent lattice, with total size $N=7200$. Inset: zoom into the flat band at $\varepsilon=0$. No effect of $\delta W_O$ is seen on any of the flat bands except for the junction flat band ($|\varepsilon|>2$), while for $\delta W_v$ all flat bands are broadened.
}
    \label{fig7}
\end{figure}

For site disorder, the on-site energies $\varepsilon_v$ in Eq.~(\ref{eq:H}) are chosen independently from a box distribution. We consider two variants. In the first, disorder is confined to the original sites, $v\in O$, with width $\delta W_O$. In the second, disorder is applied to all sites, $v\in V$, with width $\delta W_V$. As shown in Fig.~\ref{fig7}, these two cases affect the flat bands qualitatively differently.

When disorder is confined to the original sites, the chain-induced flat bands remain unaffected. This follows from the destructive-interference condition in Eq.~(\ref{eq:interference}), which ensures that these eigenstates have support only on the chain sites and therefore do not couple to the disordered original sites. In contrast, the junction-induced flat bands are strongly affected by disorder on the original sites, since these states have substantial weight on the junction sites. Finally, when on-site disorder is applied to all sites, all flat bands are strongly perturbed.

\subsection{Random magnetic flux}

Another form of disorder arises from random magnetic fluxes threading the lattice. This disorder is conceptually different because it explicitly breaks time-reversal symmetry.  The random-flux model is physically justified as an effective tight-binding description in which disorder enters through random phases on hopping amplitudes, representing spatially fluctuating magnetic fields. It describes systems where electrons acquire random Aharonov–Bohm phases from inhomogeneous magnetic flux \cite{Sugiyama1993,Avishai1993,Aronov1994,Furusaki1999},  emergent gauge fields \cite{Furusaki1999}, or length of bonds in an optical network \cite{berkovits2024}. Random flux is incorporated into Eq.~(\ref{eq:H}) through Peierls phases,
\[
t_{uv}=t\,e^{i\theta_{uv}},
\]
where each $\theta_{uv}$ is chosen independently from a box distribution on $[0,2\pi]$.

\begin{figure}
    \includegraphics[width=0.5\textwidth]{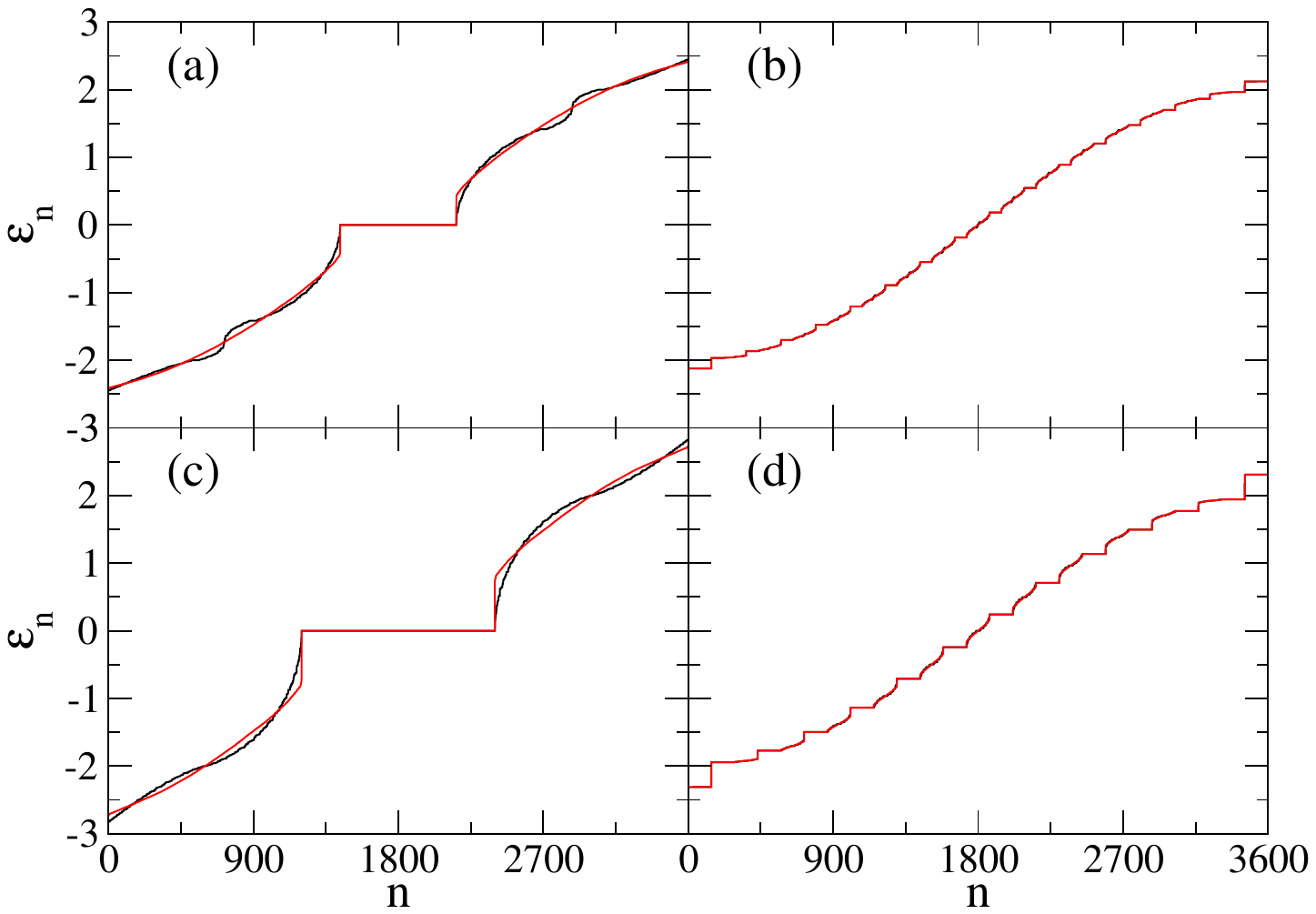}
    \caption{The $n$-th energy level $\varepsilon_n$ as a function of level index $n$ for super$^{L}$honeycomb and Lieb-$L$ lattices in the presence of random magnetic flux. (a) A single realization of a superhoneycomb lattice generated from a $40\times36$ honeycomb parent lattice and inflated once, with total size $N=3600$. The hopping amplitudes acquire random Peierls phases. (b)--(d) The same type of flux disorder for single realizations of: (b) a super$^{16}$honeycomb lattice generated from a $12\times12$ honeycomb parent lattice; (c) a Lieb-$11$ lattice generated from a $40\times30$ square parent lattice; and (d) a Lieb-$12$ lattice generated from a $12\times12$ square parent lattice. All cases contain $N=3600$ sites. In all cases the flat bands are unaffected by the random magnetic flux, while the Dirac cones are eliminated.}
    \label{fig8}
\end{figure}

States supported only on the chains effectively form decoupled one-dimensional subsystems and are therefore insensitive to random magnetic phases, since these phases can be gauged away along each chain. Consequently, the chain-induced flat bands remain unaffected by random magnetic flux, as observed in Fig.~\ref{fig8}. Junction states, which are localized primarily on original sites, are likewise only weakly perturbed by the magnetic field. In contrast, the Dirac cones lose their characteristic $\sqrt{n}$ behavior in the vicinity of the flat band, and a hard gap opens; see, for example, Fig.~\ref{fig8}(a) and Fig.~\ref{fig8}(c).

\section{Random  bond inflation}

\begin{figure*}
\centering
    \includegraphics[width=0.8\textwidth]{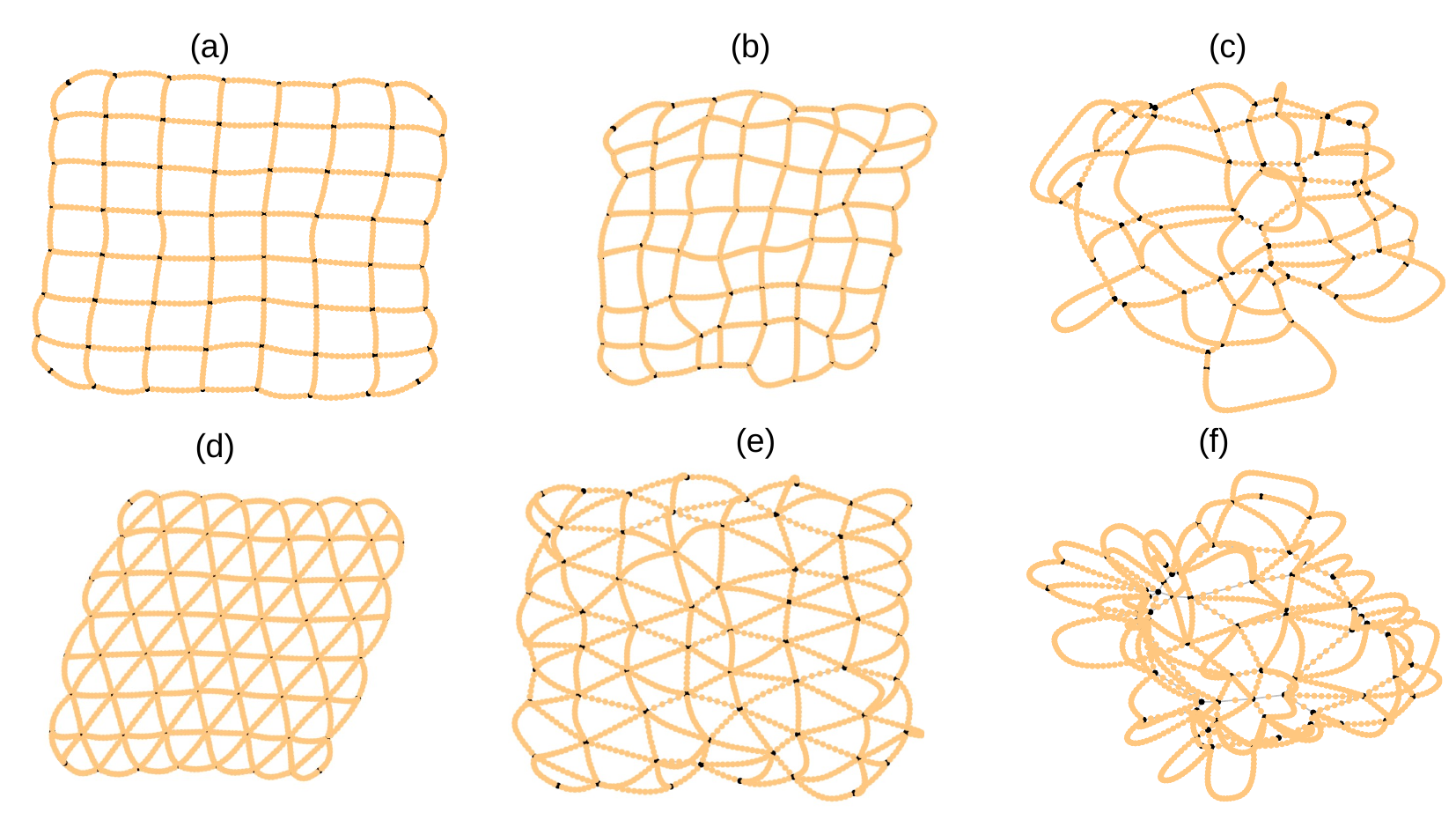}
    \caption{Segments of  bond-inflated Lieb and triangular lattices. (a) A Lieb-$10$ lattice generated from an $8\times8$ square parent lattice, where each original  bond is uniformly inflated into a chain of length $L=10$. (b) A lattice obtained by applying the same total number of inflation steps as in (a), but inflating only original  bonds chosen at random at each step. (c) A lattice obtained by applying the same total number of inflation steps as in (a), but allowing any  bond to be chosen at random for inflation at each step. (d) As in (a), but for a super$^{10}$triangular lattice generated from an $8\times8$ triangular parent lattice. (e,f) The triangular-lattice analogs of (b,c), respectively.}
    \label{fig9}
\end{figure*}

We now turn to disorder generated directly by the inflation process. Instead of inflating every original bond exactly $L$ times, we consider random protocols in which one repeatedly selects an  bond and inflates it by one step, for a total of $K$ inflation steps. The resulting structure is generally a random graph with average chain length
\[
\bar{L} = \frac{K}{N_E^{(0)}} + 1,
\]
where $N_E^{(0)}$ is the number of  bonds in the parent lattice. Such a graph retains the large-scale connectivity of the parent lattice but lacks translational symmetry; see Fig.~\ref{fig1}(e,f) and Fig.~\ref{fig9}.

We consider two random inflation protocols. In the first, only original  bonds may be selected. For $N_E^{(0)}$ original  bonds and $K$ inflation steps, the probability of obtaining a chain of length $L$ is
\begin{equation}
P_1(L) =
\binom{K}{L-1}
\left(\frac{1}{N_E^{(0)}}\right)^{L-1}
\left(1-\frac{1}{N_E^{(0)}}\right)^{K-L+1},
\end{equation}
where  $L=1,\dots,K+1$. For large $N_E^{(0)}$ and $K$, this approaches a Poisson distribution,
\begin{equation}
P_1(L) \to e^{-K/N_E^{(0)}}\,
\frac{\left(K/N_E^{(0)}\right)^{L-1}}{(L-1)!},
\qquad L=1,2,\dots.
\end{equation}

In the second protocol, any  bond in the graph may be chosen at each inflation step. This is a P{\'o}lya urn process \cite{eggenberger1923,mahmoud2008}: already long chains are more likely to be selected again. In this case,
\begin{equation}
P_2(L)=
\frac{\binom{N_E^{(0)}+K-L-1}{N_E^{(0)}-2}}
{\binom{N_E^{(0)}+K-1}{N_E^{(0)}-1}},
\qquad L=1,\dots,K+1.
\end{equation}
For large $N_E^{(0)}$ and $K$,
\begin{equation}
P_2(L)=
\frac{N_E^{(0)}}{K+N_E^{(0)}}
\left(\frac{K}{K+N_E^{(0)}}\right)^{L-1},
\qquad L=1,2,\dots.
\end{equation}
This distribution is much broader than the Poisson case.

\begin{figure}
    \includegraphics[width=0.5\textwidth]{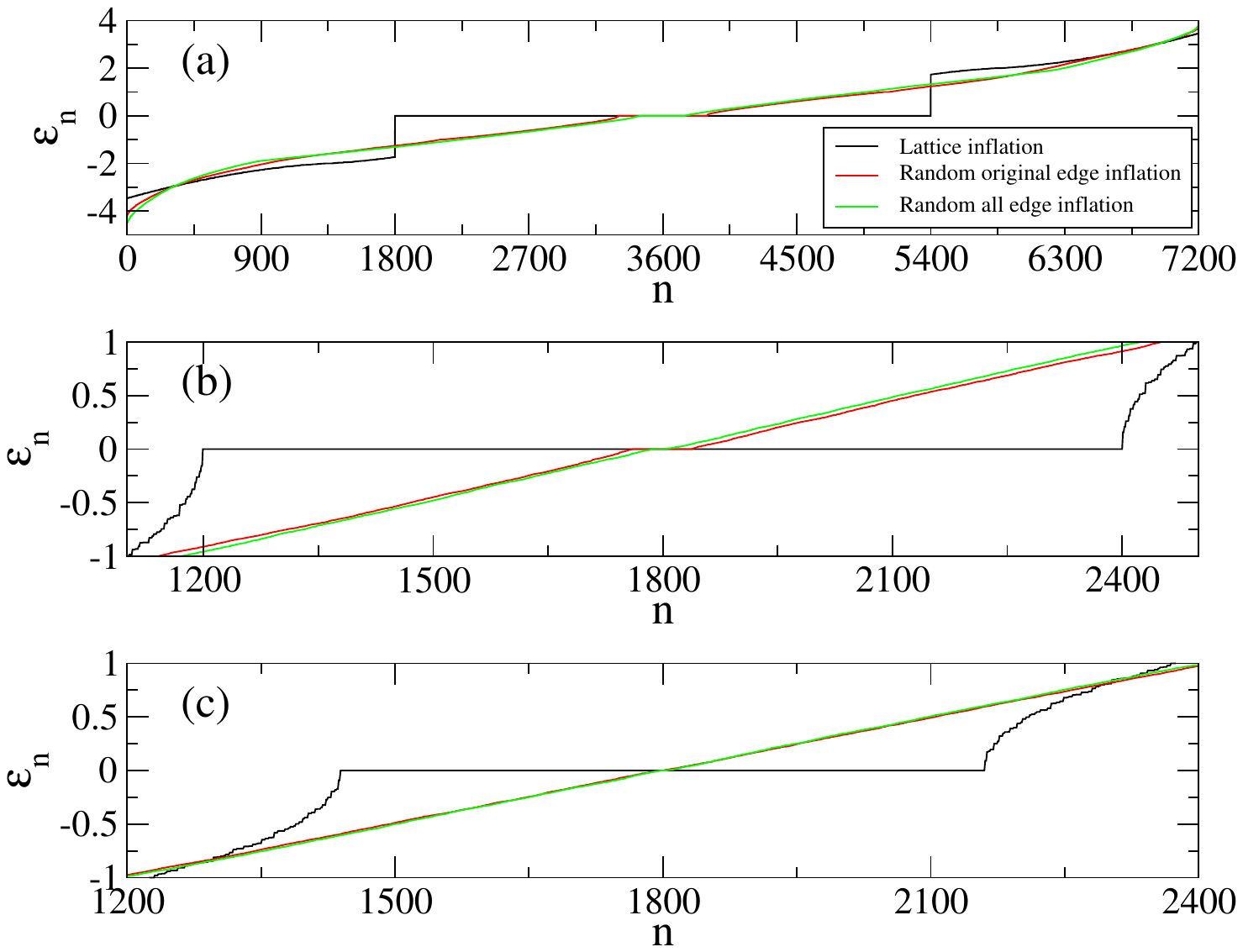}
    \caption{The $n$-th energy level $\varepsilon_n$ as a function of the level index $n$ for (a) super$^{1}$triangular ($N=7200$, $N_O=50\times36=1800$), (b) Lieb-$1$ ($N=3600$, $N_O=40\times30=1200$), and (c) super$^{1}$honeycomb ($N=3600$, $N_O=40\times36=1440$) lattices. In each case, three curves are shown: the ordered lattice with uniform chain length $L=1$ (black), random inflation applied only to original  bonds (red), and random inflation in which any  bond may be inflated (green). In the random cases, the total number of inflation steps is $K=N-N_O$. For both random inflation protocols, the degeneracy of the flat band at $\varepsilon=0$ decreases but remains substantial for the inflated triangular and square lattices, while it vanishes for the inflated honeycomb lattice.}
    \label{fig10}
\end{figure}

\begin{figure}
    \includegraphics[width=0.5\textwidth]{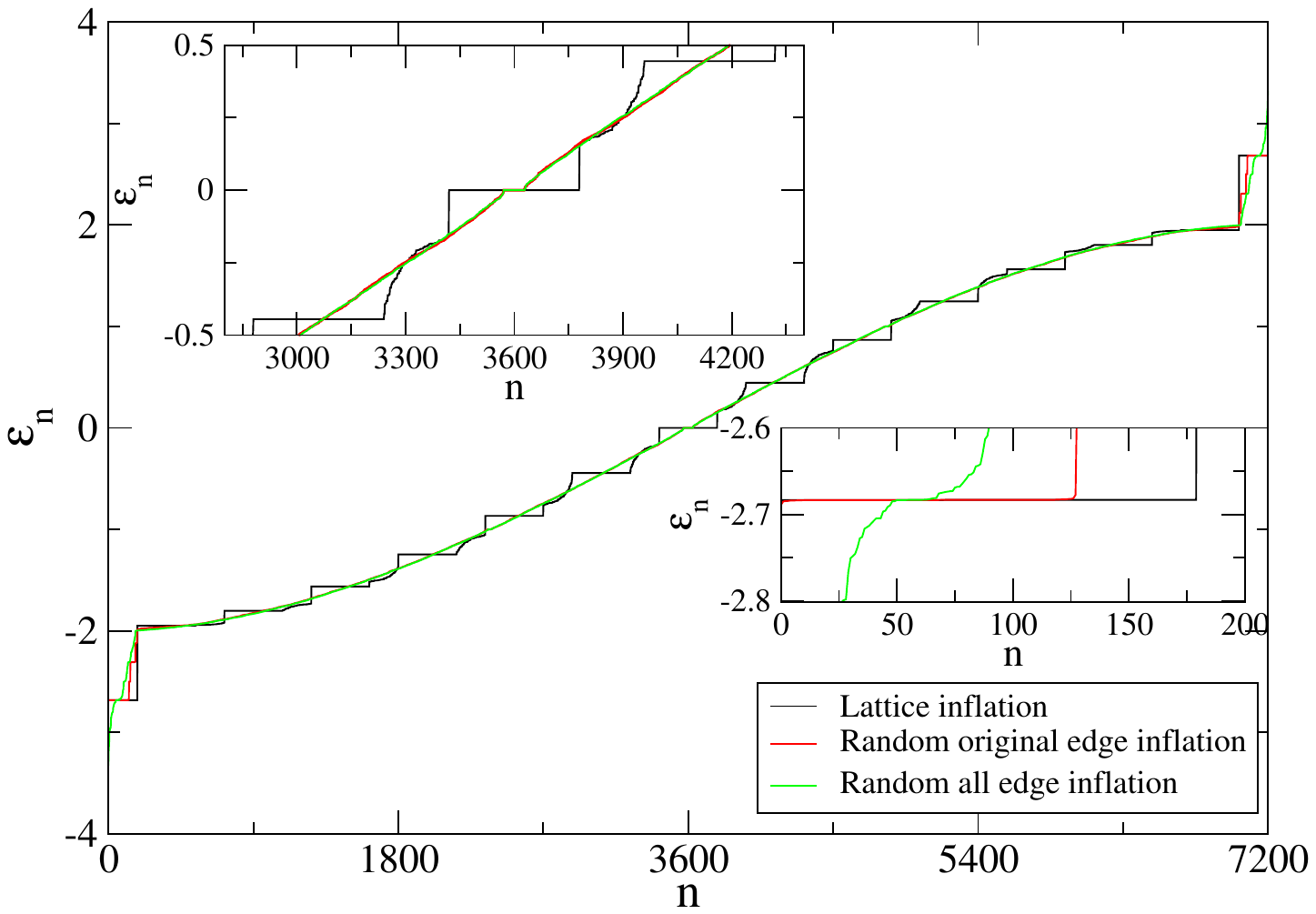}
    \caption{The $n$-th energy level $\varepsilon_n$ as a function of the level index $n$ for a super$^{13}$triangular lattice with $N=7200$ and $N_O=18\times10=180$. As in Fig.~\ref{fig10}, the three curves correspond to the ordered lattice (black), random inflation applied only to the original  bonds (red), and random inflation in which any  bond may be inflated (green). In the random cases, the total number of inflation steps is $K=N-N_O$. Top inset: zoom into the vicinity of $\varepsilon=0$. Bottom inset: zoom into the spectral edge. The two random inflation protocols yield nearly identical results for $|\varepsilon|<2$, while significant differences appear in the degeneracy of the junction-induced flat band.}
    \label{fig11}
\end{figure}

In Figs.~\ref{fig10} and \ref{fig11}, we compare $\varepsilon_n$ for three cases: (i) the ordered lattice, in which all chains have the same length $L$; (ii) random inflation applied only to the original  bonds; and (iii) random inflation in which any  bond may be inflated. In the random cases, the number of inflation steps is $K=N-N_O$.

As shown in Figs.~\ref{fig1} and \ref{fig9}, the distribution of chain lengths differs significantly between the two random inflation protocols, ranging from Poisson-like to broad, effectively power-law behavior. Remarkably, this difference leads to quantitative changes but does not alter the qualitative appearance of the flat bands.

Since the random graphs are generally not bipartite and do not appear to obey any special symmetry, it is surprising that, under suitable conditions, two spectral regions may still exhibit flat-band features. The first is at $\varepsilon=0$. The second occurs near the spectral edges, for $|\varepsilon|>2$. Notably, these flat-band states persist even in the presence of random magnetic flux, indicating that they are not the result of fine-tuned interference alone.

The origins of these robust flat-band features are different. The flat band at $\varepsilon=0$ is essentially combinatorial and arises from the structure of the random graph $G=(V,E)$ and its adjacency matrix $A$. In contrast, the flat bands at $|\varepsilon|>2$ originate from junctions to which chains are attached whose lengths are significantly larger than the broadening localization length $\xi$.

\subsection{Zero-energy flat band}

We first consider the origin of the flat band at $\varepsilon=0$. For this purpose, we recall some graph-theoretic notions \cite{Godsil2001,Cvetkovic2010}. A matching in a graph $G$ is a set of edges no two of which share a common vertex, and a maximum matching is a matching of largest possible size, whose cardinality (number of elements contained within that set) we denote by $\nu(G)$. The multiplicity of the zero eigenvalue of the adjacency matrix $A$, i.e., the nullity
\[
\eta(G)=\dim\ker A,
\]
is closely related to this combinatorial quantity. For trees (or sets of trees known as a forest) one has the exact identity
\begin{equation}
\eta(G)=N-2\nu(G),
\label{eq:eta}
\end{equation}
whereas for graphs with loops this relation generally ceases to be exact.

A common source of mismatch arises from applying tree-based intuition to graphs that contain loops. On a tree, the constraints imposed by $A\psi=0$ propagate without closure: each vertex introduces an independent linear condition, and one obtains the exact relation in Eq.~(\ref{eq:eta}). Physically, amplitudes can be assigned recursively along branches, and unmatched vertices correspond to genuinely independent zero-energy degrees of freedom.

In contrast, when the graph contains loops, these constraints are no longer independent. The condition $A\psi=0$ around a loop imposes a consistency relation: amplitudes propagated along different paths must agree when the paths meet again. This leads to additional global constraints. As a result, some candidate zero modes inferred from local tree-like reasoning are removed, while others may appear.

A simple illustration is provided by the cycle graph $C_J$, i.e., a loop of $J$ vertices. For $J=3$, one has $\nu(C_3)=1$, so $N-2\nu(C_3)=1$, whereas $\eta(C_3)=0$. For $J=4$, one has $\nu(C_4)=2$, so $N-2\nu(C_4)=0$, whereas $\eta(C_4)=2$. In general,
\begin{equation}
\nu(C_J)=\left\lfloor \frac{J}{2}\right\rfloor,
\qquad
\eta(C_J)=
\begin{cases}
2, & J\equiv 0 \pmod 4,\\
0, & J\not\equiv 0 \pmod 4.
\end{cases}
\end{equation}
Since $N=J$, it follows that
\begin{equation}
J-2\nu(C_J)=
\begin{cases}
0, & J \text{ even},\\
1, & J \text{ odd}.
\end{cases}
\end{equation}
Thus, cycle graphs explicitly demonstrate that the relation between the nullity $\eta(G)$ and the matching deficiency $N-2\nu(G)$ depends sensitively on the parity and structure of loops. Beyond trees and forests, no universal equality holds.

\begin{figure}
    \includegraphics[width=0.5\textwidth]{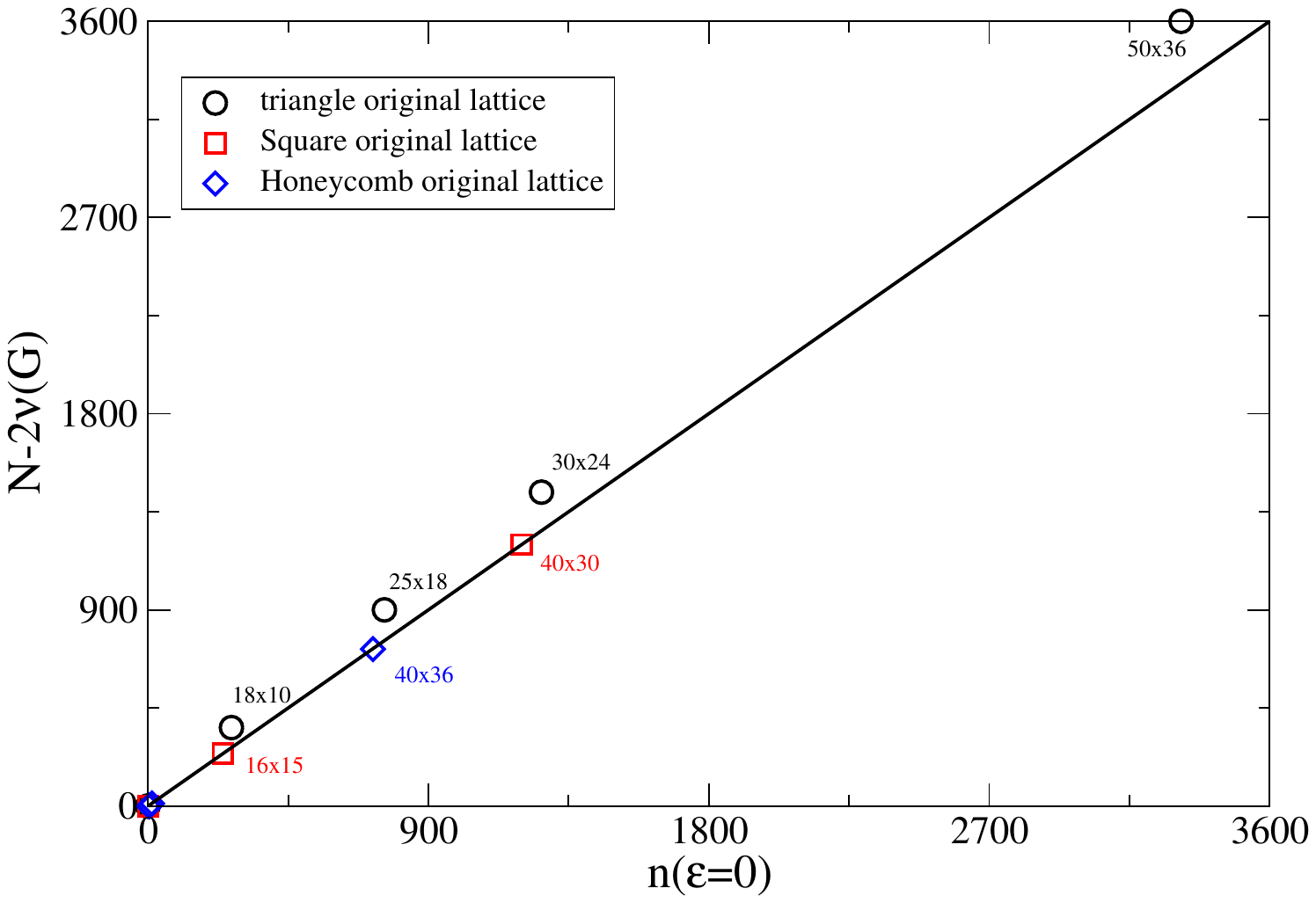}
    \caption{The number of zero-energy states, $n(\varepsilon=0)$, as a function of the matching-deficiency estimate $N-2\nu(G)$ for various ordered  bond-inflated lattices. Shown are inflated lattices of size $N=7200$ derived from triangular parent lattices (black circles), lattices of size $N=3600$ derived from square parent lattices (red squares), and lattices of size $N=3600$ derived from honeycomb parent lattices (blue diamonds). Points away from the vicinity of $(0,0)$ are labeled by the sizes of the parent lattices.}
    \label{fig12}
\end{figure}

Despite the presence of loops on all scales, Eq.~(\ref{eq:eta}) remains useful as an estimate for the number of zero-energy states in ordered  bond-inflated lattices. Fig. \ref{fig12} compares $n(\varepsilon=0)$ with the matching-deficiency estimate $N-2\nu(G)$ for a broad range of triangular, square, and honeycomb parent lattices.

For even $L$, there are no zero-energy states, or at most $O(1)$ states in the inflated honeycomb case, so $n(\varepsilon=0)\approx 0$. In all of these cases one also finds $N-2\nu(G)=0$. For odd $L$, there is a substantial number of zero-energy states. In these cases, the quantity $N-2\nu(G)$ provides a reasonable estimate of $n(\varepsilon=0)$: it is nearly exact for the inflated square and honeycomb lattices, while for the inflated triangular lattices it shows noticeable but still moderate deviations.

\begin{figure}
    \includegraphics[width=0.5\textwidth]{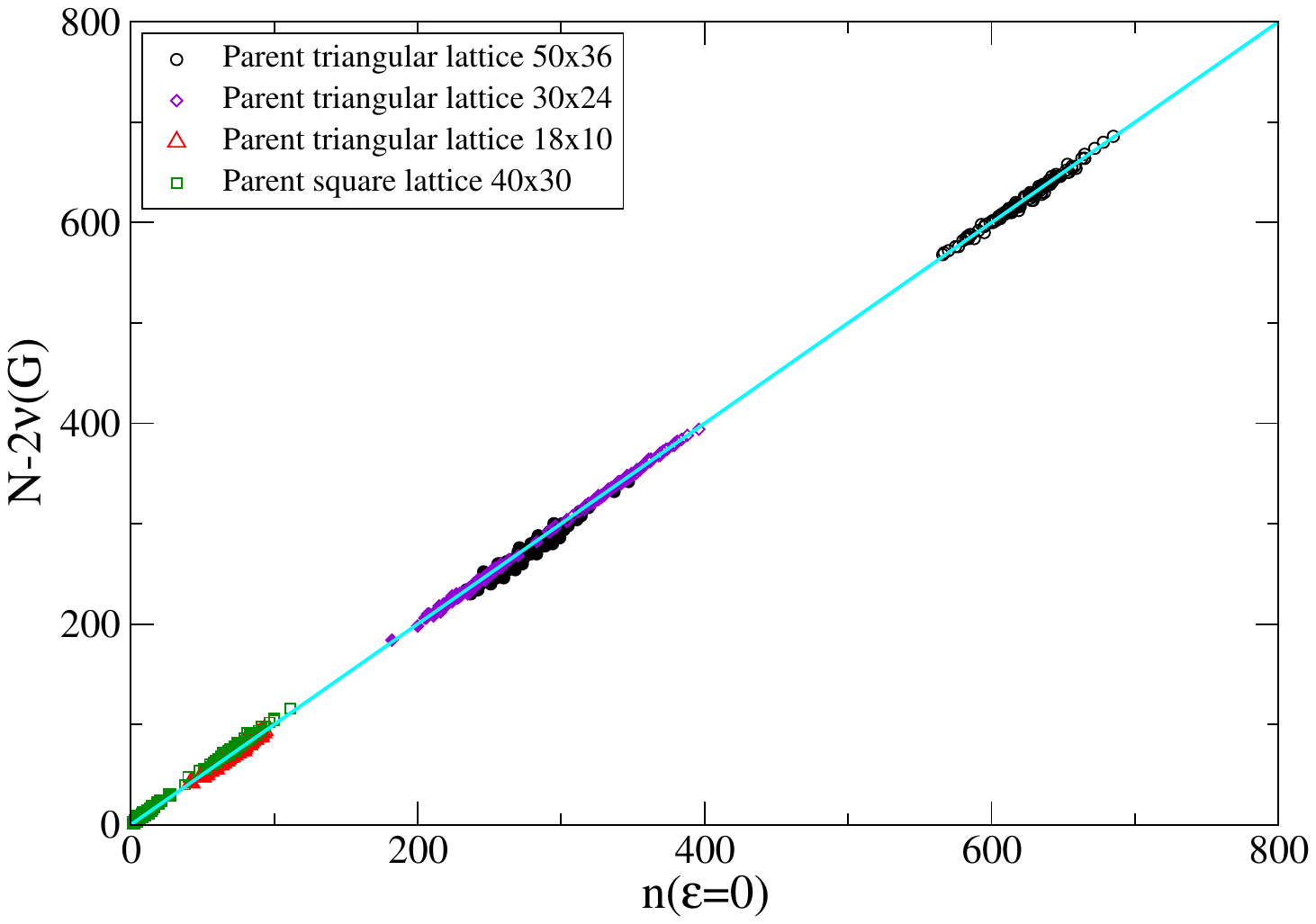}
    \caption{
    The number of zero-energy states, $n(\varepsilon=0)$, as a function of the matching-deficiency estimate $N-2\nu(G)$ for $100$ random realizations generated from several parent lattices. Shown are systems of size $N=7200$ derived from triangular parent lattices and systems of size $N=3600$ derived from a square parent lattice. For each parent lattice, both random inflation protocols are included: inflation applied only to original  bonds (empty symbols) and inflation applied to any  bond (filled symbols).}
    \label{fig13}
\end{figure}

Once random inflation is applied to a regular lattice seed, the resulting structures share the same large-scale topology, namely the same loop skeleton, although the number of vertices along each loop fluctuates. Somewhat surprisingly, the relation between $n(\varepsilon=0)$ and the matching prediction $N-2\nu(G)$ remains close to equality for each random realization, despite potentially large fluctuations in $n(\varepsilon=0)$ itself.

This behavior is clearly demonstrated in Fig.~\ref{fig13}, where $100$ random realizations are generated from triangular and square parent lattices. For each realization, $n(\varepsilon=0)$ is compared with $N-2\nu(G)$, and the agreement is remarkably accurate, with most data points lying close to the line $n(\varepsilon=0)=N-2\nu(G)$. Honeycomb lattices are not shown because, as seen in Fig.~\ref{fig10}, for these parent lattices both $n(\varepsilon=0)$ and $N-2\nu(G)$ are approximately zero.

Thus, the matching deficiency provides a useful estimate of the number of zero-energy eigenstates in randomly  bond-inflated lattices. Although these graphs contain numerous loops, they retain a locally tree-like structure. It is therefore plausible that the contributions from different random loops self-average, leading to the observation that, for a given realization, the relation $n(\varepsilon=0)\approx N-2\nu(G)$ holds much better than in the corresponding ordered lattices.

\subsection{Junction flat band}

We now turn to the origin of the robust flat-band features at $|\varepsilon|>2$. These arise from junctions to which chains are attached whose lengths exceed the broadening localization length $\xi$. For the triangular parent lattice, the relevant junction energy is
\[
\varepsilon = \pm\sqrt{\frac{k^2}{k-1}} = \pm\sqrt{\frac{36}{5}} \approx \pm2.68,
\]
with $k=6$; see Eq.~(\ref{eq:junc_e}).

We first consider the protocol in which only original  bonds are inflated. Let $\bar{L}=K/(kN_O)+1$ denote the average chain length after $K$ inflation steps. The probability that a given chain has length at least $\xi$ is
\begin{equation}
\lim_{N_O\to\infty}{\tilde P}_1(L\ge \xi)
=
1-e^{-\bar{L}}\sum_{j=0}^{\xi-1}\frac{\bar{L}^{j}}{j!}.
\end{equation}
For the random realization corresponding to Fig.~\ref{fig11}, with $N_O=180$, $K=7020$, $\bar{L}=13$, and $\xi=9$, one obtains ${\tilde P}_1(L\ge \xi)\approx 0.924$. Thus, the probability that all six chains attached to a junction exceed $\xi$ is approximately $({\tilde P}_1(L\ge \xi))^6  \approx 0.62$,
and the expected number of states in the junction flat band is $N_{\mathrm{Junc}} \approx 0.62 N_O \approx 110$. This is close to the value observed numerically. Averaging over $200$ realizations yields $\langle N_{\mathrm{Junc}}\rangle = 113.48 \pm 4.48$, in excellent agreement with this estimate.

For the second protocol, in which any  bond may be selected at each inflation step, the probability that a chain reaches length at least $\xi$ is
\begin{equation}
  \lim_{N_O\to\infty} {\tilde P}_2(L\ge \xi)
  =
\left(\frac{\bar{L}-1}{\bar{L}}\right)^{\xi-2},
\end{equation}
For the same parameters, this yields ${\tilde P}_2(L\ge \xi)=(12/13)^7 \approx 0.57$. The probability that all six chains attached to a junction exceed $\xi$ is then $({\tilde P}_2(L\ge \xi))^6 \approx 0.03$, which leads to the estimate $N_{\mathrm{Junc}} \approx 0.03\,N_O \approx 6$. This value is lower than the numerical result by roughly a factor of three. Averaging over $200$ random realizations gives $\langle N_{\mathrm{Junc}}\rangle = 17.49 \pm 3.42$.

The origin of this discrepancy lies in the much broader fluctuations in chain length generated by the second protocol. Short chains remove some junctions from the ideal flat-band condition due to hybredization, but they also reduce resonance overlap with neighboring junctions, thereby weakening spectral broadening and effectively shortening the localization region. A more complete treatment incorporating these competing effects remains an open problem. The numerical results nevertheless suggest that the net effect is to increase the number of states in the junction flat band relative to the simplest estimate.

\section{Discussion and Conclusions}

In this paper we studied the spectral and localization properties of lattices generated by repeated  bond inflation of square, honeycomb, and triangular parent lattices. This construction provides a simple and flexible route for producing both periodic decorated lattices and random graphs while preserving the large-scale connectivity of the parent lattice. Despite this geometric simplicity, the resulting spectra display a rich hierarchy of flat bands whose origin depends on symmetry, local interference constraints, and junction-induced localization.

For the ordered  bond-inflated lattices, three distinct mechanisms were identified. The first is the appearance of chain-induced flat bands at the eigenenergies of the finite one-dimensional chains replacing the original  bonds. These bands arise because chain eigenmodes can be combined so that the amplitudes vanish on the original junction sites, thereby preventing hybridization between neighboring motifs. The second mechanism is the zero-energy flat band, which appears in bipartite  bond-inflated lattices with sublattice imbalance and is protected by chiral symmetry. The third mechanism is the formation of junction bands near the spectral edges, associated with exponentially localized states centered on the original sites when the attached chains are sufficiently long. Although these three classes of flat bands originate from different physical considerations, they can all be understood as consequences of local rank deficiency or orthogonality to the coupling subspace that links a local motif to the extended lattice.

A central result of this paper is that flat-band physics survives, to a surprising extent, even when translational symmetry is destroyed by random  bond inflation. In the two random protocols considered here, the chain-length distribution changes qualitatively, ranging from Poisson-like to broad, power-law behavior. Nevertheless, the spectral signatures of flat bands remain visible. In particular, a substantial accumulation of states persists at zero energy and, for sufficiently large average chain length, a second concentration of states emerges near the energies associated with junction localization. The persistence of these features under random magnetic flux further demonstrates that they are not merely a consequence of fine-tuned phase interference, but instead reflect robust structural properties of the inflated graphs.

The zero-energy sector of the random  bond-inflated lattices is especially noteworthy. Although these graphs typically contain loops on many scales and need not remain bipartite, the number of zero modes is found to track the matching-deficiency estimate $N-2\nu(G)$ with unexpectedly high accuracy for individual realizations. This behavior is much closer than one would generally expect for graphs with loops, where the exact tree identity $\eta(G)=N-2\nu(G)$ no longer applies. Our results suggest that the random inflated graphs remain sufficiently locally tree-like that loop corrections largely self-average, leaving the matching deficiency as an effective predictor of the nullity. This observation points to an interesting connection between spectral graph theory and flat-band formation in spatially embedded random networks.

The junction flat bands in the random systems reveal a complementary aspect of this robustness. Their existence depends primarily on the probability that all chains connected to a given junction exceed the broadening localization length $\xi$. For the protocol in which only original  bonds are inflated, this probability is well captured by a Poisson description and leads to quantitative agreement with the number of observed junction states. For the protocol in which any  bond can be inflated, the broader chain-length distribution produces stronger fluctuations and a less accurate estimate, although the qualitative mechanism remains the same. The residual discrepancy likely reflects the competition between two effects: short chains suppress the existence of junction-localized states at a given vertex, but at the same time they reduce hybridization and broadening with neighboring junctions. A more refined theory that incorporates these competing effects remains an open problem.

Several extensions of the present work would be worthwhile. First, it would be interesting to develop an analytical theory for the zero-mode statistics of random inflated graphs beyond the matching-deficiency approximation, explicitly incorporating the role of loop parity and local cycle structure. Second, the interplay between flat bands and interactions in these inflated geometries may lead to unusual correlated phases, particularly in the presence of robust zero-energy degeneracies. Third, since the construction is purely graph based, it should be straightforward to generalize it to other parent lattices, higher dimensions, quasiperiodic tilings, or networks with designed degree distributions. Finally, the robustness of the flat bands under magnetic flux and several forms of disorder suggests that  bond-inflated lattices may provide a useful platform for experimental realizations in photonic, electric-circuit, phononic, and magnonic systems.

In summary,  bond inflation generates a broad family of lattices and random graphs in which flat bands arise from three complementary mechanisms: chain interference, chiral sublattice imbalance, and junction localization. These mechanisms remain operative far beyond the conventional periodic flat-band setting and survive substantial structural randomness. The results presented here therefore establish  bond-inflated lattices as a natural framework for studying localization, spectral singularities, and robust flat-band phenomena in both ordered and disordered graph-based systems.

\begin{acknowledgments}
  I thank D. Gutman, S. Carr, P. Matveeva and S. Havlin, for insightful discussions that initiated this research.
\end{acknowledgments}

\bibliography{einf3}

@article{Sutherland1986,
  author  = {Bill Sutherland},
  title   = {Localization of electronic wave functions due to local topology},
  journal = {Phys. Rev. B},
  volume  = {34},
  pages   = {5208--5211},
  year    = {1986},
  doi     = {10.1103/PhysRevB.34.5208}
}

@article{Lieb1989,
  author  = {Lieb, Elliott H.},
  title   = {Two theorems on the {H}ubbard model},
  journal = {Phys. Rev. Lett.},
  volume  = {62},
  number  = {10},
  pages   = {1201--1204},
  year    = {1989},
  doi     = {10.1103/PhysRevLett.62.1201}
}

@article{Mielke1991,
  author  = {Mielke, Andreas},
  title   = {Ferromagnetism in the {H}ubbard model on line graphs and further considerations},
  journal = {J. Phys. A: Math. Gen. },
  volume  = {24},
  pages   = {3311--3321},
  year    = {1991}
}

@article{Tasaki1992,
  author  = {Tasaki, Hal},
  title   = {Ferromagnetism in the {H}ubbard models with degenerate single-electron ground states},
  journal = {Phys. Rev. Lett.},
  volume  = {69},
  number  = {10},
  pages   = {1608--1611},
  year    = {1992},
  doi     = {10.1103/PhysRevLett.69.1608}
}

@article{Bergman2008,
  author  = {Bergman, Doron L. and Wu, Congjun and Balents, Leon},
  title   = {Band touching from real-space topology in frustrated hopping models},
  journal = {Phys. Rev. B},
  volume  = {78},
  pages   = {125104},
  year    = {2008},
  doi     = {10.1103/PhysRevB.78.125104}
}

@article{Flach2014,
  author  = {Flach, Sergej and Leykam, Daniel and Bodyfelt, Joshua D. and Matthies, Peter and Desyatnikov, Anton S.},
  title   = {Detangling flat bands into {F}ano lattices},
  journal = {Europhys. Lett.},
  volume  = {105},
  number  = {3},
  pages   = {30001},
  year    = {2014},
  doi     = {10.1209/0295-5075/105/30001}
}

@article{Leykam2018,
  author  = {Leykam, Daniel and Andreanov, Alexei and Flach, Sergej},
  title   = {Artificial flat band systems: from lattice models to experiments},
  journal = {Adv. Phys. X},
  volume  = {3},
  number  = {1},
  pages   = {1473052},
  year    = {2018},
  doi     = {10.1080/23746149.2018.1473052}
}

@article{CastroNeto2009,
  author  = {Castro Neto, A. H. and Guinea, F. and Peres, N. M. R. and Novoselov, K. S. and Geim, A. K.},
  title   = {The electronic properties of graphene},
  journal = {Rev.s of Modern Physics},
  volume  = {81},
  pages   = {109--162},
  year    = {2009},
  doi     = {10.1103/RevModPhys.81.109}
}

@article{Derzhko2015,
  author  = {Derzhko, Oleg and Richter, Johannes and Maksymenko, Mykola},
  title   = {Strongly correlated flat-band systems: The route from {H}eisenberg spins to {H}ubbard electrons},
  journal = {International Journal of Modern Physics B},
  volume  = {29},
  number  = {23},
  pages   = {1530007},
  year    = {2015},
  doi     = {10.1142/S0217979215300078}
}

@book{Godsil2001,
  author    = {Godsil, Chris and Royle, Gordon},
  title     = {Algebraic Graph Theory},
  publisher = {Springer},
  year      = {2001},
  series    = {Graduate Texts in Mathematics},
  volume    = {207},
  doi       = {10.1007/978-1-4613-0163-9}
}

@book{Cvetkovic1995,
  author    = {Cvetkovi{\'c}, Drago{\v{s}} M. and Doob, Michael and Sachs, Horst},
  title     = {Spectra of Graphs: Theory and Application},
  edition   = {3},
  publisher = {Johann Ambrosius Barth Verlag},
  address   = {Heidelberg--Leipzig},
  year      = {1995}
}

@article{Bhattacharya2019,
  title        = {Flat bands and nontrivial topological properties in an extended Lieb lattice},
  author       = {Bhattacharya, Ankita and Pal, Biplab},
  journal      = {Phys. Rev. B},
  volume       = {100},
  pages        = {235145},
  year         = {2019},
  doi          = {10.1103/PhysRevB.100.235145},
}

@article{Xia2018,
  title   = {Unconventional Flatband Line States in Photonic Lieb Lattices},
  author  = {Xia, Shiqi and Ramachandran, Ajith and Xia, Shiqiang and Li, Denghui and Liu, Xiuying and Tang, Liqin and Hu, Yi and Song, Daohong and Xu, Jingjun},
  journal = {Phys. Rev. Lett.},
  volume  = {121},
  pages   = {263902},
  year    = {2018},
  doi     = {10.1103/PhysRevLett.121.263902}
}

@article{Slot2017,
  title        = {Experimental realization and characterization of an electronic {L}ieb lattice},
  author       = {Slot, Marlou R. and Gardenier, Thomas S. and Jacobse, Peter H. and van Miert, Guido C. P. and Kempkes, Sander N. and Zevenhuizen, Stephan J. M. and Morais Smith, Cristiane and Vanmaekelbergh, Daniel and Swart, Ingmar},
  journal      = {Nature Physics},
  volume       = {13},
  pages        = {672--676},
  year         = {2017},
  doi          = {10.1038/nphys4105},
}

@article{Zhang2017,
  title        = {New edge-centered photonic square lattices with flat bands},
  author       = {Zhang, Da and Zhang, Yiqi and Zhong, Hua and Li, Changbiao and Zhang, Zhaoyang and Zhang, Yanpeng and Beli{\'c}, Milivoj R.},
  journal      = {Annals of Physics},
  volume       = {382},
  pages        = {160--169},
  year         = {2017},
  doi          = {10.1016/j.aop.2017.04.016},
}

@article{Hanafi2022,
  title        = {Localized dynamics arising from multiple flat bands in a decorated photonic {L}ieb lattice},
  author       = {Hanafi, H. and Menz, P. and McWilliam, A. and Imbrock, J. and Denz, C.},
  journal      = {APL Photonics},
  volume       = {7},
  pages        = {111301},
  year         = {2022},
  doi          = {10.1063/5.0109840},
}

@article{Krawczyk2023,
  title        = {Compact localized states in magnonic {L}ieb lattices},
  author       = {Krawczyk, Maciej and Gruszecki, Przemyslaw and Swiercz, Wojciech and Klos, Jaroslaw W. and Chumak, Andrii V. and Sokolovskyy, Mykhailo L. and others},
  journal      = {Scientific Reports},
  volume       = {13},
  pages        = {12676},
  year         = {2023},
  doi          = {10.1038/s41598-023-39816-w},
}

@article{Shen2010,
  author  = {Shen, R. and Shao, L. B. and Wang, Baigeng and Xing, D. Y.},
  title   = {Single {D}irac cone with a flat band touching on line-centered-square optical lattices},
  journal = {Phys. Rev. B},
  volume  = {81},
  pages   = {041410(R)},
  year    = {2010},
  doi     = {10.1103/PhysRevB.81.041410}
}

@article{Inui1994,
  author    = {M. Inui and S. A. Trugman and E. Abrahams},
  title     = {Unusual properties of midband states in systems with off-diagonal disorder},
  journal   = {Phys. Rev. B},
  volume    = {49},
  pages     = {3190--3196},
  year      = {1994},
}

@article{khatua21,
  author    = {S. Khatua and S. Srinivasan and R. Ganesh},
  title     = {State selection in frustrated magnets},
  journal   = {Phys. Rev. B},
  volume    = {103},
  pages     = {174412},
  year      = {2021},
}

@article{he23,
  author    = {E. He and R. Ganesh},
  title     = {Bound states without potentials: Localization at singularities},
  journal   = {Phys. Rev. A},
  volume    = {108},
  pages     = {022202},
  year      = {2023},
}

@article{Hasan2010,
  author  = {Hasan, M. Zahid and Kane, Charles L.},
  title   = {Colloquium: Topological insulators},
  journal = {Rev. Mod. Phys.},
  volume  = {82},
  pages   = {3045--3067},
  year    = {2010},
  doi     = {10.1103/RevModPhys.82.3045}
}

@article{Qi2011,
  author  = {Qi, Xiao-Liang and Zhang, Shou-Cheng},
  title   = {Topological insulators and superconductors},
  journal = {Rev. Mod. Phys.},
  volume  = {83},
  pages   = {1057--1110},
  year    = {2011},
  doi     = {10.1103/RevModPhys.83.1057}
}

@article{Polini2013,
  author  = {Polini, Marco and Guinea, Francisco and Lewenstein, Maciej and Manoharan, Hari C. and Pellegrini, Vittorio},
  title   = {Artificial honeycomb lattices for electrons, atoms and photons},
  journal = {Nat. Nanotechnol,},
  volume  = {8},
  pages   = {625--633},
  year    = {2013},
  doi     = {10.1038/nnano.2013.161}
}

@article{eggenberger1923,
  author    = {Eggenberger, F. and P{\'o}lya, George},
  title     = {{\"U}ber die Statistik verketteter Vorg{\"a}nge},
  journal   = {Z. Angew. Math. Mech.},
  volume    = {3},
  number    = {4},
  pages     = {279--289},
  year      = {1923}
}

@book{mahmoud2008,
  author    = {Mahmoud, Hosam M.},
  title     = {P{\'o}lya Urn Models},
  publisher = {CRC Press},
  year      = {2008}
}

@book{Cvetkovic2010,
  author    = {Drago{\v{s}} Cvetkovi{\'c} and Peter Rowlinson and Slobodan Simi{\'c}},
  title     = {An Introduction to the Theory of Graph Spectra},
  publisher = {Cambridge University Press},
  year      = {2010}
}

@article{Apaja2010,
  author  = {V. Apaja and M. Hyrk{\"a}s and M. Manninen},
  title   = {Flat bands, Dirac cones, and atom dynamics in an optical lattice},
  journal = {Phys. Rev. A},
  volume  = {82},
  pages   = {041402},
  year    = {2010},
  doi     = {10.1103/PhysRevA.82.041402}
}

@article{Weeks2010,
  author  = {C. Weeks and M. Franz},
  title   = {Topological insulators on the Lieb and perovskite lattices},
  journal = {Phys. Rev. B},
  volume  = {82},
  pages   = {085310},
  year    = {2010},
  doi     = {10.1103/PhysRevB.82.085310}
}

@article{Nita2013,
  author  = {M. Nita and B. Ostahie and A. Aldea},
  title   = {Spectral and transport properties of the Lieb lattice},
  journal = {Phys. Rev. B},
  volume  = {87},
  pages   = {125428},
  year    = {2013},
  doi     = {10.1103/PhysRevB.87.125428}
}

@article{Diebel2016,
  author  = {F. Diebel and D. Leykam and S. Kroesen and C. Denz and A. S. Desyatnikov},
  title   = {Conical diffraction and composite Lieb bosons in photonic lattices},
  journal = {Phys. Rev. Lett.},
  volume  = {116},
  pages   = {183902},
  year    = {2016},
  doi     = {10.1103/PhysRevLett.116.183902}
}

@article{Danieli2024,
  author  = {Carlo Danieli and Alexei Andreanov and Daniel Leykam and Sergej Flach},
  title   = {Flat band fine-tuning and its photonic applications},
  journal = {Nanophotonics},
  volume  = {13},
  number  = {21},
  pages   = {3925--3944},
  year    = {2024},
  doi     = {10.1515/nanoph-2024-0135},
}

@article{Leykam2017,
  author  = {Daniel Leykam and Joshua D. Bodyfelt and Anton S. Desyatnikov and Sergej Flach},
  title   = {Localization of weakly disordered flat-band states},
  journal = {Eur. Phys. J. B},
  volume  = {90},
  pages   = {1},
  year    = {2017},
  doi     = {10.1140/epjb/e2016-70586-9},
  eprint  = {1601.03784},
  archivePrefix = {arXiv}
}

@article{Danieli2015,
  author  = {Carlo Danieli and Joshua D. Bodyfelt and Sergej Flach},
  title   = {Flat-band engineering of mobility edges},
  journal = {Phys. Rev. B},
  volume  = {91},
  pages   = {235134},
  year    = {2015},
  doi     = {10.1103/PhysRevB.91.235134},
  eprint  = {1502.06690},
  archivePrefix = {arXiv}
}

@article{Cadez2021,
  author  = {Tilen \v{C}ade\v{z} and Yeongjun Kim and Alexei Andreanov and Sergej Flach},
  title   = {Metal-insulator transition in infinitesimally weakly disordered flatbands},
  journal = {Phys. Rev. B},
  volume  = {104},
  pages   = {L180201},
  year    = {2021},
  doi     = {10.1103/PhysRevB.104.L180201},
  eprint  = {2107.11365},
  archivePrefix = {arXiv}
}

@article{Kim2023,
  author  = {Yeongjun Kim and Tilen \v{C}ade\v{z} and Alexei Andreanov and Sergej Flach},
  title   = {Flat-band induced metal-insulator transitions for weak magnetic flux and spin-orbit disorder},
  journal = {Phys. Rev. B},
  volume  = {107},
  pages   = {174202},
  year    = {2023},
  doi     = {10.1103/PhysRevB.107.174202},
  eprint  = {2211.09410},
  archivePrefix = {arXiv}
}

@article{Chalker2010,
  author = {Chalker, J. T. and Pickles, T. S. and Shukla, P.},
  title = {Anderson localization in tight-binding models with flat bands},
  journal = {Phys. Rev. B},
  volume = {82},
  pages = {104209},
  year = {2010},
  doi = {10.1103/PhysRevB.82.104209}
}

@article{Leykam2013,
  author = {D. Leykam, S. Flach, Omri Bahat-Treidel and A. S. Desyatnikov},
  title = {Flat band states: Disorder and nonlinearity},
  journal = {Phys. Rev. B},
  volume = {88},
  pages = {224203},
  year = {2013},
  doi = {10.1103/PhysRevB.88.224203}
}

@article{Shukla2019,
  author = {Shukla, Pragya},
  title = {Disorder perturbed flat bands II: Density of states and criticality},
  journal = {Phys. Rev. B},
  volume = {99},
  pages = {054206},
  year = {2019},
  doi = {10.1103/PhysRevB.99.054206}
}

@article{Rosen2025,
  author = {Rosen, Ilan T. and Muschinske, Sarah and Barrett, Cora N. and Rower, David A. and Das, Rabindra and Kim, David K. and Niedzielski, Bethany M. and Schuldt, Meghan and Serniak, Kyle and Schwartz, Mollie E. and Yoder, Jonilyn L. and Grover, Jeffrey A. and Oliver, William D.},
  title = {Flat-Band (De)localization Emulated with a Superconducting Qubit Array},
  journal = {Physical Review X},
  volume = {15},
  number = {2},
  pages = {021091},
  year = {2025},
  doi = {10.1103/PhysRevX.15.021091},
  eprint = {2410.07878},
  archivePrefix = {arXiv}
}

@article{Dresselhaus2025,
  author = {Dresselhaus, Elizabeth J. and Avdoshkin, Alexander and Jia, Zhetao and Secli, Matteo and Kante, Boubacar and Moore, Joel E.},
  title = {A tale of two localizations: coexistence of flat bands and Anderson localization in a photonics-inspired amorphous system},
  journal = {Physical Review B},
  volume = {112},
  pages = {064202},
  year = {2025},
  doi = {10.1103/q8vz-b83f},
  eprint = {2404.17578},
  archivePrefix = {arXiv}
}

@article{Liu2022,
  author  = {Jie Liu and Carlo Danieli and Jianxin Zhong and Rudolf A. R{\"o}mer},
  title   = {Unconventional delocalization in a family of three-dimensional Lieb lattices},
  journal = {Phys. Rev. B},
  volume  = {106},
  pages   = {214204},
  year    = {2022},
  doi     = {10.1103/PhysRevB.106.214204},
  url     = {https://doi.org/10.1103/PhysRevB.106.214204}
}

@article{Liu2021,
  author  = {Jie Liu and Xiaoyu Mao and Jianxin Zhong and Rudolf A. R{\"o}mer},
  title   = {Localization properties in Lieb lattices and their extensions},
  journal = {Annals of Physics},
  volume  = {435},
  pages   = {168544},
  year    = {2021},
  doi     = {10.1016/j.aop.2021.168544}
}

@article{Danieli2024a,
  author  = {Carlo Danieli and Jie Liu and Rudolf A. R{\"o}mer},
  title   = {Quantum engineering for compactly localized states in disordered Lieb lattices},
  journal = {Eur. Phys. J. B},
  volume  = {97},
  pages   = {128},
  year    = {2024},
  doi     = {10.1140/epjb/s10051-024-00745-w},
  url     = {https://doi.org/10.1140/epjb/s10051-024-00745-w}
}

@article{Marques2023,
  author  = {A. M. Marques and J. M{\"o}gerle and G. Pelegr{\'i} and S. Flannigan and R. G. Dias and A. J. Daley},
  title   = {Kaleidoscopes of Hofstadter butterflies and Aharonov-Bohm caging from $2n$-root topology in decorated square lattices},
  journal = {Phys. Rev. Research},
  volume  = {5},
  pages   = {023110},
  year    = {2023},
  doi     = {10.1103/PhysRevResearch.5.023110}
}

@article{Sugiyama1993,
  author    = {Sugiyama, Taishi and Nagaosa, Naoto},
  title     = {Localization in a random magnetic field in 2D},
  journal   = {Physical Review Letters},
  volume    = {70},
  number    = {13},
  pages     = {1980--1983},
  year      = {1993},
  doi       = {10.1103/PhysRevLett.70.1980},
  publisher = {American Physical Society}
}

@article{Avishai1993,
  author    = {Avishai, Y. and Hatsugai, Y. and Kohmoto, M.},
  title     = {Localization problem of a two-dimensional lattice in a random magnetic field},
  journal   = {Physical Review B},
  volume    = {47},
  number    = {15},
  pages     = {9561--9565},
  year      = {1993},
  doi       = {10.1103/PhysRevB.47.9561},
  publisher = {American Physical Society}
}

@article{Aronov1994,
  author    = {Aronov, A. G. and Mirlin, A. D. and W{\"o}lfle, P.},
  title     = {Localization of charged quantum particles in a static random magnetic field},
  journal   = {Physical Review B},
  volume    = {49},
  number    = {23},
  pages     = {16609--16629},
  year      = {1994},
  doi       = {10.1103/PhysRevB.49.16609},
  publisher = {American Physical Society}
}

@article{Furusaki1999,
  author        = {Furusaki, Akira},
  title         = {Anderson localization due to a random magnetic field in two dimensions},
  journal       = {Physical Review Letters},
  volume        = {82},
  number        = {3},
  pages         = {604--607},
  year          = {1999},
  doi           = {10.1103/PhysRevLett.82.604},
  eprint        = {cond-mat/9808059},
  archivePrefix = {arXiv},
  primaryClass  = {cond-mat.mes-hall},
  publisher     = {American Physical Society}
}

@article{Berkovits2024,
  author    = {Berkovits, Richard},
  title     = {Inducing a metal-insulator transition through systematic alterations of local rewriting rules in a quantum graph},
  journal   = {Physical Review B},
  volume    = {109},
  pages     = {L220201},
  year      = {2024},
  month     = jun,
  doi       = {10.1103/PhysRevB.109.L220201},
  publisher = {American Physical Society}
}

@article{vidal1998,
  author  = {J. Vidal, R. Mosseri and B. Dou\c{c}ot},
  title   = {Aharonov-Bohm Cages in Two-Dimensional Structures},
  journal = {Phys. Rev. Lett.},
  volume  = {81},
  pages   = {5888--5891},
  year    = {1998},
  doi     = {10.1103/PhysRevLett.81.5888}
}

@article{vidal2001,
  author  = {j. Vidal, P. Butaud, B. Dou\c{c}ot and R. Mosseri},
  title   = {Disorder and interactions in Aharonov-Bohm cages},
  journal = {Phys. Rev. B},
  volume  = {64},
  pages   = {155306},
  year    = {2001},
  doi     = {10.1103/PhysRevB.64.155306}
}

\appendix

\section{Line graphs and flat bands}
\label{app_line}

To clarify the appearance of flat bands in line graphs, consider a simple undirected graph $G=(V,E)$ with $|V|=N$ vertices and $|E|=M$ edges. The unoriented vertex-edge incidence matrix $B$ of size $N\times M$ is defined by
\begin{equation}
B_{v,e} =
\begin{cases}
1, & \text{if vertex } v \text{ is an endpoint of edge } e,\\
0, & \text{otherwise}.
\end{cases}
\end{equation}
Each column of $B$ contains exactly two nonzero entries. A standard identity of algebraic graph theory \cite{Godsil2001,Cvetkovic1995} states that
\begin{equation}
B^{\mathsf T}B = A(L(G)) + 2I,
\end{equation}
where $L(G)$ is the line graph of $G$, whose vertices correspond to edges of $G$, and where two vertices in $L(G)$ are adjacent if the corresponding edges in $G$ share a common endpoint. Thus, the tight-binding Hamiltonian on a line graph can be written, up to a constant shift, as
\begin{equation}
H_{L(G)} = B^{\mathsf T}B - 2I.
\end{equation}
Because $B^{\mathsf T}B$ is positive semidefinite, the kernel of $H_{L(G)}$ coincides with the kernel of $B$,
\begin{equation}
\ker H_{L(G)} = \ker B.
\end{equation}
The dimension of the flat-band subspace is therefore
\begin{equation}
\dim\ker B = M - \rank(B).
\end{equation}
Since generically $M>N$, one obtains $\dim\ker B>0$, leading to a substantial flat band at $\varepsilon=0$. Flat-band eigenstates are characterized by the local constraint
\begin{equation}
(B\psi)_v = \sum_{e\ni v}\psi_e = 0
\qquad \forall v\in V,
\end{equation}
which expresses destructive interference at every original vertex. This algebraic structure provides a rigorous origin for flat bands in broad families of line-graph lattices \cite{Mielke1991,Bergman2008}.

\section{Junction states}
\label{app_junction}

Consider a junction site $o$ to which $k$ chains are attached, obtained, for example, by repeated edge inflation as in Fig.~\ref{fig2}. In the limit of infinitely long chains, $L\to\infty$, the ground and highest excited states can be obtained using an ansatz wave function \cite{khatua21,he23}. Using the notation of Fig.~\ref{fig2}(iii), we take
\begin{eqnarray}
\phi_o &=& C, \\
\psi^{(c)}_1 &=& \pm e^{-\alpha}\phi_o, \\
\psi^{(c)}_p &=& \pm e^{-\alpha}\psi^{(c)}_{p-1}, \qquad p>1,
\end{eqnarray}
where $C$ is a normalization constant, and the $\pm$ sign corresponds to the ground state ($+$) and the highest excited state ($-$), respectively.

Substituting this ansatz into the tight-binding Schr\"odinger equation yields
\begin{eqnarray}
\varepsilon \phi_o &=& -t \sum_{c=1}^{k}\psi^{(c)}_1, \\
\varepsilon \psi^{(c)}_p &=& -t\left(\psi^{(c)}_{p-1}+\psi^{(c)}_{p+1}\right).
\end{eqnarray}
For the highest excited state one may equivalently replace $t$ by $-t$.

By symmetry, the wave functions on different chains are identical, $\psi^{(c)}_p=\psi_p$, which reduces the equations to
\begin{eqnarray}
\varepsilon \phi_o &=& -kt\,\psi_1, \\
\varepsilon \psi_p &=& -t\left(\psi_{p-1}+\psi_{p+1}\right).
\end{eqnarray}
Solving these equations yields
\[
\frac{\varepsilon}{t} = \pm \sqrt{\frac{k^2}{k-1}},
\qquad
\alpha = \frac{\ln(k-1)}{2}.
\]
These solutions lie below and above the one-dimensional chain band, respectively. The localization length is
\[
\zeta = \frac{1}{\alpha} = \frac{2}{\ln(k-1)}.
\]
Consequently, the solution is insensitive to boundary conditions at the ends of the chains, and the localized state remains robust provided that the chains attached to the junction are longer than several lattice spacings.

\end{document}